\documentclass[sn-mathphys,Numbered]{sn-jnl}

\usepackage{graphicx}%
\usepackage{multirow}%
\usepackage{amsmath,amssymb,amsfonts}%
\usepackage{amsthm}%
\usepackage{mathrsfs}%
\usepackage[title]{appendix}%
\usepackage{xcolor}%
\usepackage{textcomp}%
\usepackage{manyfoot}%
\usepackage{booktabs}%
\usepackage{algorithm}%
\usepackage{algorithmicx}%
\usepackage{algpseudocode}%
\usepackage{listings}%
\usepackage{braket}%
\usepackage{soul}%
\usepackage{subfig}%
\usepackage{multicol}%
\usepackage{array,multirow}

\newtheorem{definition}{Definition}%

\begin{document}

\title[Nav-Q: Quantum Deep Reinforcement Learning for Collision-Free Navigation of Self-Driving Cars
]{Nav-Q: Quantum Deep Reinforcement Learning for Collision-Free Navigation of Self-Driving Cars \vspace{-.5em}}



\author[1,2]{\fnm{Akash} \sur{Sinha}}

\author[2]{\fnm{Antonio} \sur{Macaluso}}\email{antonio.macaluso@dfki.de}

\author[2]{\fnm{Matthias} \sur{Klusch}}

\affil[1]{\orgdiv{Department of Mathematics and Computer Science}, \orgname{Saarland University}, \orgaddress{ \city{Saarbruecken}, \state{Saarland}, \country{Germany}}}

\affil[2]{\orgdiv{Agents and Simulated Reality Department}, \orgname{German Research Center for Artificial Intelligence (DFKI)}, \orgaddress{\city{Saarbruecken}, \state{Saarland}, \country{Germany}}}



\abstract{
The task of collision-free navigation (CFN) of self-driving cars is an NP-hard problem usually tackled using Deep Reinforcement Learning (DRL). While DRL methods have proven to be effective, their implementation requires substantial computing resources and extended training periods to develop a robust agent.
On the other hand, quantum reinforcement learning has recently demonstrated faster convergence and improved stability in simple, non-real-world environments. 

In this work, we propose Nav-Q, the first quantum-supported DRL algorithm for CFN of self-driving cars, that leverages quantum computation for improving the training performance without the requirement for onboard quantum hardware. Nav-Q is based on the actor-critic approach, where the critic is implemented using a hybrid quantum-classical algorithm suitable for near-term quantum devices.
We assess the performance of Nav-Q using the CARLA driving simulator, a de facto standard benchmark for evaluating state-of-the-art DRL methods. 
Our empirical evaluations showcase that Nav-Q surpasses its classical counterpart in terms of training stability and, in certain instances, with respect to the convergence rate. 
Furthermore, we assess Nav-Q in relation to effective dimension, unveiling that the incorporation of a quantum component results in a model with greater descriptive power compared to classical baselines.
Finally, we evaluate the performance of Nav-Q using noisy quantum simulation, observing that the quantum noise deteriorates the training performances but enhances the exploratory tendencies of the agent during training.
 }




\maketitle

\section{Introduction}\label{sec1}




The pursuit of autonomous navigation for self-driving cars has ushered in a new era of transportation, promising safer and more efficient journeys. Central to this endeavor is the challenge of collision-free navigation (CFN), a complex optimization problem that lies at the heart of ensuring the safety and reliability of autonomous vehicles. Over the years, significant strides have been made in tackling this problem, with researchers and engineers harnessing the power of computational methods to chart paths that avoid obstacles and minimize risks \cite{Zhu_2022, chib2023recent, RL_Survey3}.

In the domain of autonomous navigation, a prevalent approach involves conceptualizing the problem as Partially Observable Markov Decision Processes (POMDPs) \cite{hyleap, gupta2022hybrid} which poses a significant computational challenge when dealing with real-world scenarios. The current state-of-the-art solutions rely on Deep Reinforcement Learning (DRL) \cite{Zhu_2022, chib2023recent, RL_Survey3}, which can be integrated with POMDP planning \cite{hyleap, gupta2022hybrid}. While these techniques have demonstrated promising results, they demand substantial computing resources and training time to develop a robust reinforcement learning (RL) agent.

In parallel, the field of quantum reinforcement learning (QRL) has recently emerged as an innovative paradigm that harnesses the computational capabilities of quantum computing in conjunction with the principles of reinforcement learning (RL). Early experiments \cite{jerbi2021parametrized, lan2021variational, chen2020variational, Skolik_2022} have showcased the potential of QRL methods to accelerate training times when applied to simplified, non-real-world environments, often exemplified by the benchmark domains provided by OpenAI Gym \cite{brockman2016openai}. However, a critical gap remains to be bridged as none of these methods have been tested in the challenging, real-world environment as the one of CFN of self-driving cars. 
Furthermore, many of the suggested methodologies presuppose the substitution of traditional neural networks, which provide action directives, with parametrized quantum circuits (PQCs). In the context of CFN, this implies the requirement of a quantum computing device installed onboard the vehicle for experimentation—a requirement that is impractical to fulfill.

This paper delves into the feasibility and potential advantages of quantum-supported DRL within the domain of CFN of self-driving cars. Specifically, we introduce a novel quantum-supported architecture that leverages quantum computation during training, resulting in state-of-the-art performance during testing.
Precisely, the contributions are the following:  
\begin{enumerate}
    \item Introducing Nav-Q, the first quantum-supported Deep Reinforcement Learning (DRL) approach that combines quantum and classical neural networks to tackle the CFN problem of self-driving cars. Nav-Q employs an actor-critic algorithm, with the critic implemented through a hybrid quantum-classical approach, enhancing the performance during training without requiring an onboard quantum device for testing.
    \item Extend the Data Reuploading technique \cite{P_rez_Salinas_2020} by proposing a new encoding strategy for quantum embedding named Qubit Independent Data Encoding and Processing (QIDEP), allowing for the versatile encoding of high-dimensional classical data while providing flexibility in the number of qubits used.

\item
Train and evaluate Nav-Q on the CARLA-CTS1.0 benchmark \cite{gupta2022hybrid}, a widely recognized standard for testing DRL models in the context of self-driving cars. Nav-Q's performance is compared with its natural classical counterpart, NavA2C (Section \ref{sec:NavA2C}), which exclusively utilizes classical computational resources.
The comprehensive results indicate that Nav-Q demonstrates improved stability across various initializations of the architecture and, in some instances, a faster convergence rate. 
Additionally, Nav-Q attains state-of-the-art performance in the driving policy while substantially reducing the number of parameters. It also showcases a heightened potential for learnability when evaluating effective dimension as a metric to assess the model's ability to learn complex patterns from data.

    \item Train Nav-Q on a simulated noisy quantum system with the parameter-shift rule to estimate the expected performance when experiments are performed on near-term quantum hardware. 

\end{enumerate}

 The rest of the paper is structured as follows. A discussion of the related works, quantum and classical, is reported in Sect. \ref{sec: related work}. This is followed by the formalization of the CFN problem as a POMDP and a detailed description of an advanced classical solution named NavA2C, which serves as the classical baseline, in Section \ref{sec:problem}. Our quantum-supported solution Nav-Q is described in Sect. \ref{sec:Methods} and its performance comparatively evaluated in Sect. \ref{sec:Experiments}. Section \ref{sec:Conclusions} concludes with a summary of achievements and future work.

\section{Related Works}\label{sec: related work}

In this section, we provide a summary of the existing literature relevant to two key aspects of this paper's contributions. The first part delves into current state-of-the-art approaches for addressing the CFN problem, primarily centered around DRL algorithms. Secondly, we explore the QRL landscape, which focuses on training parameterized quantum circuits to function as agents within classical environments.

\subsection{Deep Reinforcement Learning for Collision-Free Navigation}\label{sec:DRL for CFN}

The state-of-the-art approach for addressing the CFN problem of self-driving cars typically involves formulating it as a POMDP and solving it through DRL. Different approaches are possible depending on the specific problem the DRL algorithm is tasked with, ranging from simple lane following to complex tasks like end-to-end driving in urban scenarios \cite{Zhu_2022, chib2023recent, RL_Survey3}.
These methods exhibit variations in the DRL algorithm used, system architecture, reward design, tasks, input, and actions performed by the agent.

Recently, different actor-critic-based architectures for DRL have been proposed. In particular, A3C approaches \cite{A2C,mirowski2017learning} execute on multiple parallel instances of the environment to use the
parallel actor-learners for stabilizing effect on training
process. 
The adaptation of this approach in the context of CFN is referred to as GA3C-CADRL \cite{Everett_2021} and has been tested for mobile robot navigation. In this case, the observation of the environment is not presented as a picture image but rather as the position, velocity, and orientation of the robot. Also, an LSTM \cite{LSTM} is used to encode temporal and spatial data in this scenario. Alternatively, A2C is the synchronous version of the A3C model, that waits for each agent to finish its experience before conducting an update. In this respect, 
A2C-CADRL-p has been recently introduced \cite{gupta2022hybrid} as a synchronous variant of GA3C-CADRL that utilizes the output of path planning and the perceived environment as input to the network, as opposed to end-to-end learning, enabling the agent to navigate the environment with a specific goal.

Beyond A2C and A3C approaches, whose performance are comparable, recent work has highlighted Proximal Policy Optimization (PPO) as a standout approach for self-driving cars \cite{schulman2017proximal,tang2020learning}. PPO excels due to its stability, sample efficiency, and adaptability to continuous action spaces. By preventing large policy updates, it ensures stable learning, particularly crucial for safety-critical applications. Moreover, its compatibility with parallel processing accelerates learning in data-intensive scenarios.

At the time of writing this paper, Pennylane~\cite{bergholm2022pennylane} with the PyTorch plugin did not support parameter broadcasting\footnote{\url{https://docs.pennylane.ai/en/stable/code/api/pennylane.broadcast.html}}. Consequently, the quantum version of PPO, which requires batch processing of data through quantum circuit simulations, incurred a substantial time cost on classical computers, rendering training these algorithms impractical. Therefore, for practical purposes, our focus is on the classical baseline NavA2C (see Section~\ref{sec:NavA2C}) based on A2C-CADRL-p \cite{gupta2022hybrid}. This approach combines a symbolic path planner and a modified A2C agent in one architecture, with the DRL agent solely responsible for the speed action. Nevertheless, the Nav-Q methodology (described in Section \ref{sec:Methods}) utilizes an actor-critic model and can be seamlessly adapted to the PPO approach, provided that the available quantum software tools support efficient parameter broadcasting.

\subsection{Quantum Reinforcement Learning}
The standard approach in Quantum Reinforcement Learning (QRL) consists of replacing a classical Neural Network with a Parametrized Quantum Circuit (PQC) to enhance performance in terms of trainability, stability, and learning policy.

Among the various alternatives, the use of a Quantum Q-Learning algorithm has been investigated and compared to a Deep Q-Network, revealing a parameter space complexity of PQCs scaling as $\mathcal{O}(N)$, in contrast to the $\mathcal{O}(N^{2})$ complexity of classical DQN \cite{chen2020variational}. This investigation focused on PQCs operating in a straightforward deterministic environment with a single input value, where both the observation space and output space were restricted to the same number of quantum bits. In a subsequent study \cite{lockwood2020reinforcement}, alternative PQC architectures and encoding schemes were explored, extending the algorithm's testing to slightly more complex environments. Another advancement involved a novel data encoding scheme using trainable weights, enhancing Quantum Q-Learning performance \cite{Skolik_2022}.
In a different vein, a Quantum Soft Actor-Critic algorithm has been proposed \cite{lan2021variational}, where the actor was implemented as a PQC, and the critic utilized a classical neural network. Another study proposed the Softmax-PQC, employing a softmax policy to successfully address various standard RL tasks based on REINFORCE \cite{reinforce, jerbi2021parametrized}. Additionally, a QRL algorithm based on PPO \cite{schulman2017proximal} has been introduced, with the actor implemented using a PQC and the critic with a classical neural network \cite{cartpole, kwak2021introduction}. Lastly, the use of Q-SAC to control a simulated robotic arm with continuous action and state spaces was demonstrated \cite{acuto2022variational}.

\begin{table}[]
\begin{tabular}{lllll}
\toprule
\textbf{Environment} & \textbf{\begin{tabular}[c]{@{}l@{}}Observation \\ Space\end{tabular}}                                                & \textbf{\begin{tabular}[c]{@{}l@{}}Observation \\ Shape\end{tabular}} & \textbf{Action Space}                                                      & \textbf{\begin{tabular}[c]{@{}l@{}}Action \\ Shape\end{tabular}} \\ 
\midrule
Acrobot  & Continuous  & (6,) & Discrete(3) & (1,)                                                             \\ 

BlackJack  &   \begin{tabular}[c]{@{}l@{}}Tuple(Discrete(32), \\ Discrete(11), \\ Discrete(2))\end{tabular} & (3,) & Discrete(2)   & (1,)                                                             \\ 

CartPole &   Continuous  &   (4,)  & Discrete(2)  & (1,)                                                             \\ 

FrozenLake   &   Discrete(16)   &   (1,)   & Discrete(4)    & (1,)                                                             \\ 

MountainCar    &   Continuous     &  (2,)  & Discrete(3)    & (1,)                                                             \\

Pendulum   & Continuous  & (3,)   & Continuous   & (1,)                                                             \\

\botrule
\end{tabular}
\caption{An overview of the characteristics of the environments used to evaluate the performance of existing quantum reinforcement learning methods.}
\label{tab:RL Environments}
\end{table}

All current QRL methods, as summarized in Table \ref{tab:RL Environments}, have been assessed in the context of OpenAI Gym environments. These environments are characterized by being static, fully observable, and featuring a low-dimensional state space. However, the CFN problem of self-driving cars is dynamic, partially observable, and involves high-dimensional observations and state spaces. As a result, existing QRL algorithms fall short in effectively addressing CFN problems.
Moreover, PQCs primarily serve to determine the actions that an agent should take within an environment. Even when adapting existing QRL methods for more intricate problems, incorporating a Quantum Processing Unit (QPU) into the car becomes a necessity in algorithm design. However, this requirement is impractical given the current challenges associated with building a real QPU.
Therefore, our goal is to introduce a novel method capable of harnessing the potential advantages of quantum computing within the realm of CFN without the need to install a QPU in cars.

\section{Problem Formulation}\label{sec:problem}
Collision-free navigation (CFN) refers to the task of driving on a drivable path from one location to another while minimizing time to goal and the number of collisions with static and dynamic obstacles. CFN task can be modeled as a POMDP \cite{hyleap, gupta2022hybrid} problem that requires a formal definition of the car model, pedestrian model, and environment. 

\subsection{Car and Pedestrian Model} \label{sec: car_ped_model}

The Car state is defined as a 4-tuple ($p^c_t$, $p^c_{goal}$, $v^c_t$, $\theta^c_t$) where $p^c_t$  $\in \mathbb{R}^2$ is the current position, $p^c_{goal}$ $\in \mathbb{R}^2$ is the goal position, $v^c_t$ $\in \mathbb{R}$ the speed and $\theta^c_t$ $\in$ [0, 2$\pi$) the orientation of the car at instance $t$. The car uses the bicycle model to approximately model the car driving on the road. We assume for the sake of this work that the car can detect all unoccluded static and dynamic obstacles within a radius of 50 m. The vehicle has a 360-degree view of its surroundings. However, in the case of dynamic obstacles, only its current position is visible; the car's perception model is unable to see the obstacle's goal position or the potential direction in which it might move. Additionally, we assume that all available sensor data is noise-free. For goal-directed navigation of the driverless car, we use the \textit{Anytime Weighted Hybrid A* Path Planner} to calculate a driveable path for the car from source to destination. This planner needs a 2-dimensional discretized cost map of the environment to plan the route. Our basic cost-map converts environmental data from the actual scene to a discretized grid map from a bird's eye view, with obstacle costs of car states defined as maximum of 1, 50, and 100, respectively, if the car is entirely on the road, partially on the sidewalk, and collides with an obstacle anywhere else. 

The pedestrian state is represented by a 4-tuple $(p_{\text{t}}^{ped}, p_{\text{goal}}^{ped}, v_{\text{t}}^{ped}, \theta_{\text{t}}^{ped})$, where  $p_{\text{t}}^{ped} \in \mathbb{R}^2$ is the current position of the pedestrian,  $p_{\text{goal}}^{ped} \in \mathbb{R}^2$ is the goal position of the pedestrian,  $v_{\text{t}}^{ped} \in \mathbb{R}$ is the speed of the pedestrian and $\theta_{\text{t}}^{ped} \in [0, 2\pi)$ is the orientation of the pedestrian at time instance $t$. We assume that the pedestrians move towards the goal in a straight line.

\subsection{Collision-free navigation problem}

The problem of CFN can be formulated as a POMDP$(S, A, T, R, \gamma, Z, O)$ as follows:



\noindent$S$: 
\hangindent=0.5cm Represents the set of states, $\{[c, \text{ped}_1, \ldots, \text{ped}_{|P|}] | c \in \mathbb{R}^2 \times \mathbb{R}^2 \times [0, 2\pi), \text{ped}_i \in \mathbb{R}^2 \times \mathbb{R}^2 \times \mathbb{R}^2 \times [0, 2\pi), 1 \leq i \leq |P| \}$, where $c$ denotes the car's state at time $t$, and $\text{ped}_i = [\text{ped}_o^i, \text{ped}_h^i]$ represents the observable and hidden state of pedestrian $i$ ($1 \leq i \leq |P|$).


\noindent $A$: 
\hangindent=0.5cm Denotes the set of all possible actions. Each action $a \in A$ is defined as a 2-tuple $(\alpha, \text{acc})$, where $\alpha$ is the steering angle and $\text{acc}$ is the speed action. The steering angles are discretized into bins of $25^\circ$ with a maximum steering angle $\alpha_{\text{max}} = 50^\circ$ in both directions. The set of all possible steering directions is $\text{Angles} = \{-50^\circ, -25^\circ, 0^\circ, 25^\circ, 50^\circ\}$. The car can choose between three distinct speed actions: $\text{SpeedActions} = \{\text{accelerate, maintain, decelerate}\}$. The accelerate action corresponds to an increase in speed by $5$ kmph, the decelerate action corresponds to a decrease in speed by $5$ kmph, and the maintain action corresponds to no change in speed.

\noindent $T$: 
\hangindent=0.5cm Represents the transition probability, $T(s_t, a_t, s_{t+1}) = p(s_{t+1}|s_t, a_t) \in [0, 1]$, which defines the likelihood of transitioning from state $s_t \in S$ to state $s_{t+1} \in S$ when taking action $a_t \in A$ at time $t$. These transitions are fully determined by car movement kinematics and the pedestrian model.

\noindent $Z$: 
\hangindent=0.5cm Signifies the set of observations, $\{\{[c, ped_1^o, ped_2^o, ..., ped_n^0]\}\}$, containing information about the car's state and the observable aspects of pedestrians in the scene.

\noindent$O$: 
\hangindent=0.5cm Denotes the observation probability, $O(s_{t+1}, a_t, o_{t+1}) = p(o_{t+1}|s_{t+1}, a_t) \in [0, 1]$, which represents the likelihood of observing $o_{t+1} \in Z$ when transitioning to $s_{t+1}$ after executing action $a_t$. 


\noindent $R$: 
\hangindent=0.5cm Stands for immediate rewards, $R(s_t, a_t) \in \mathbb{R}$, for executing action $a_t$ in state $s_t$. The reward function is formally defined in Definition \ref{def: A2C-CADRL Reward}

\noindent $\gamma$: 
\hangindent=0.5cm Represents the discount factor within the range $[0, 1]$, and we set $\gamma = 0.99$.

\subsection{Classical Baseline NavA2C} \label{sec:NavA2C}
In this section, we describe NavA2C, a DRL-based architecture that will serve as a baseline for comparison with our proposed quantum-supported method, Nav-Q. 

\begin{figure}[ht]
    \centering
     \includegraphics[width=1\linewidth]{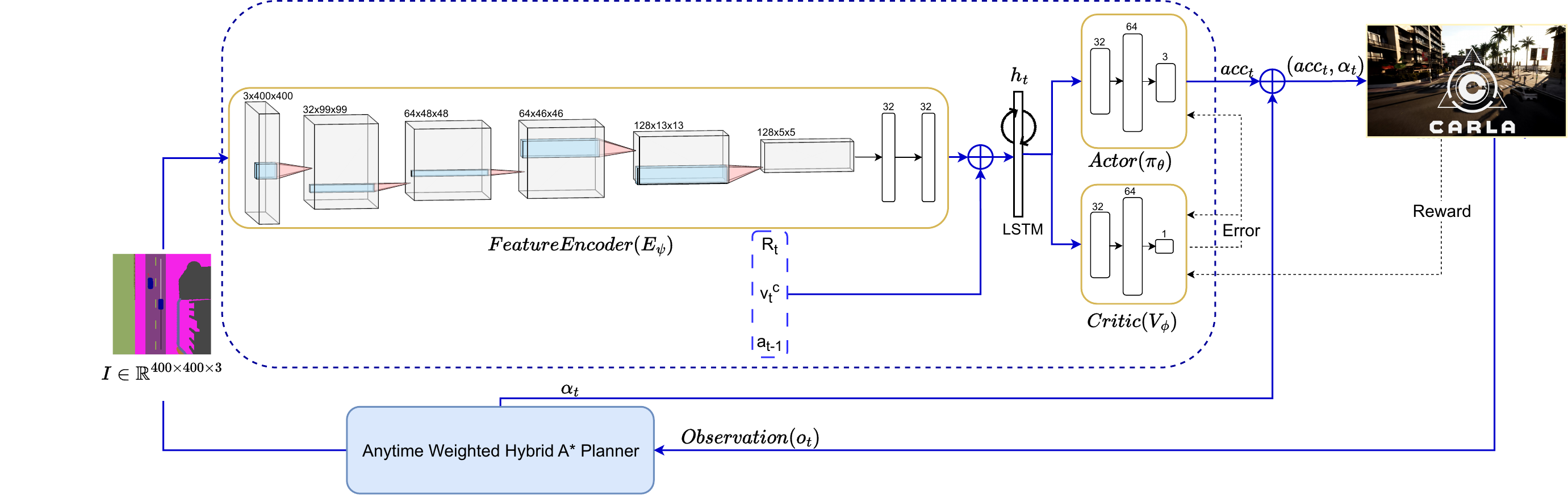}
    \caption{NavA2C architecture}
    \label{fig:NavA2C architecture}
\end{figure}

NavA2C, inspired by A2C-CADRL-p \cite{gupta2022hybrid}, combines a symbolic path planner \cite{hyleap} and a modified A2C \cite{A2C} agent, such that the DRL agent is only responsible for the speed action. The architecture is shown in Figure \ref{fig:NavA2C architecture}.

\begin{figure}[ht]
    \centering
    \includegraphics[width=0.5\linewidth]{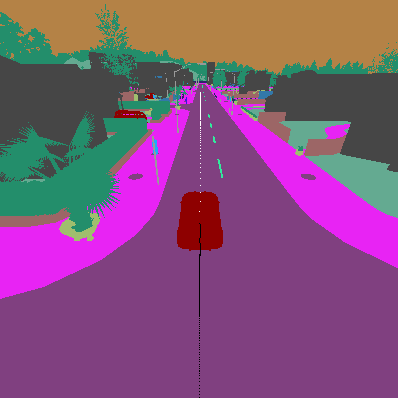}
    \caption[Car Intention]{Car Intention $ I \in \mathbb{R}^{400 \times 400 \times 3}$ is the semantic segmentation of the bird eye view of the environment. The white dotted line represents the planned path while the black dotted line represents the path already traveled by the car.  }
    \label{fig: Car Intention}
\end{figure}

The input for the DRL agent is the car intention $I$  (Figure \ref{fig: Car Intention}) which is the semantic segmentation of the environment from a bird eye view overlayed with the planned and traveled path. The DRL agent outputs the speed action $acc_t$ at every time step using car intention $I$ along with the reward $R_t$, car's velocity $v_t^c$,  and car's speed action in the previous time step $acc_{t-1}$. The steering action $\alpha_t$, needed to control the car's lateral trajectory, is derived from the path calculated by the \textit{Anytime Weighted Hybrid A* Path Planner} at every time step. The speed action $acc_t$ together with the steering angle $\alpha_t$ is passed on to CARLA simulator as a tuple $a_t = (acc_t, \alpha_t)$ which serves as the control signal for the car simulated in the CARLA environment. 

The DRL agent in NavA2C consists of four main components: Feature Encoder $E_\psi$, Actor $\pi_\theta$, LSTM, and Critic $V_\phi$. $E_\psi$ is responsible for learning a low-dimensional representation of car intention $I$ using a long sequence of CNN layers. The output of the feature encoder, $E_\psi(I)$, is concatenated with the reward $R_t \in \mathbb{R}$, the car's velocity $v_t \in \mathbb{R}^2$, and the car's speed action in the previous step $acc_{t-1} \in \mathbb{R}$. This concatenated information is passed as input to the LSTM. The LSTM's hidden state $h_{t}$ is then passed to the actor and critic networks to determine $\pi_\theta(a_t | h_t, o_t)$ and $V_\phi(h_t, o_t)$, respectively.

The Actor ($\pi_\theta$) is implemented through a dense neural network. It takes input $h_t$ and produces a $3$-dim vector with associated probabilities for speed actions (accelerate, maintain, and decelerate).
The critic $V_\phi$ is also implemented using a dense neural network and evaluates the actions chosen by the actor by estimating the expected cumulative reward associated with a given state-action pair. By leveraging these evaluations, the actor adjusts its policy to prioritize actions with higher rewards.

Specifically, the outputs of the actor and critic are the two components that allow the calculation of the loss function during training, defined as follows.
\vspace{.5em}

\begin{definition}[NavA2C Loss Function] \label{def: A2C-CADRL Loss}
    
Given the observation $o_t$, action $a_t$, value function $V_{\phi}$ with parameters $\phi$, and actor policy function $\pi_{\theta}$ with parameters $\theta$, the loss functions to train parameters $\phi$ and $\theta$ are given by,
\begin{align}
J_V(\phi) = (G_t - V_{\phi}(h_{t}, o_t))^2 
\end{align}
\begin{align}
J_{\pi}(\theta) = \log(\pi_{\theta}(a_t|h_{t}, o_t)) \cdot (G_t - V_{\phi}(h_{t}, o_t)) + \beta H(\pi_{\theta}(a_t|h_{t}, o_t))
\end{align}

\noindent where  $\beta$ is a hyperparameter that controls the weight of the entropy term,  $G_t$ is the total discounted reward, and
\begin{align}
H(\pi_{\theta}(a_t|h_{t}, o_t)) = \sum_{a_t \in \mathcal{A}} \pi_{\theta}(a_t|h_{t}, o_t) \cdot \log(\pi_{\theta}(a_t|h_{t}, o_t))
\end{align}
\noindent is the entropy of the actor policy.
\end{definition}

\vspace{.5em}

Importantly, the critic is utilized in calculating the loss function during training, providing feedback on the quality of selected actions. Its use is not required at the time of testing.
Nevertheless, opting for a critic with limited expressive power may pose challenges in capturing complex relationships, leading to slow learning, and difficulties in generalization and stability. To address these issues, selecting a critic with ample capacity to represent environmental intricacies, through advanced models, is crucial.
Thus, the primary distinction between the training architecture depicted in Figure \ref{fig:NavA2C architecture} and the corresponding testing architecture lies in the absence of the critic.

Another crucial aspect is the design of an effective reward function, as it directly impacts the learning process and the behavior of the autonomous vehicle. In fact, a meticulously crafted reward function must align the learning objectives with the desired behavior, contributing to the development of a reliable and safe autonomous driving system. Drawing inspiration from prior works \cite{Everett_2021,gupta2022hybrid}, we define the following reward function for NavA2C.\vspace{.5em}

\begin{definition}[NavA2C Reward]\label{def: A2C-CADRL Reward}
Given the current state $s_t \in S$ and previous action $a_{t-1} \in \mathbb{A}$, the reward $R_t(s_t, a_{t-1})$ is,

\[
  R_t(s_t, a_{t-1}) =
  \begin{cases}
    r_{\text{goal\_reached}} = +200
        & \begin{aligned}
            & \text{for reaching goal position, i.e., } \\
            &  p_c(t) = p_c^{\text{goal}}
        \end{aligned}          \\
        \hline
    r_{\text{hit}}=-100*\beta 
        & \begin{aligned}
           & \text{For colliding with the obstacles } \\
           & \text{(pedestrian, cars, etc.) } \beta \text{ is proportional} \\
           & \text{to the impact speed}
        \end{aligned}          \\
        \hline
    r_{\text{obstacle}}=-obstacleCost 
        & \begin{aligned}
           & \text{Penalty if there is a pedestrian or cars  } \\
           & \text{ in the hit area and }v_t^c \neq 0 \\
        \end{aligned}          \\
        \hline
    r_{\text{near\_miss}}=-10
        & \begin{aligned}
            & \text{If there is a pedestrian in the near-miss } \\
            & \text{ area of car}
        \end{aligned} \\
        \hline 
    r_{\text{over\_speeding}}=-10
        & \begin{aligned}
            & \text{For crossing the maximum speed limit of} \\
            & 50\text{ kmph}
        \end{aligned} \\
        \hline 
    r_{\text{not\_goal}}=-goal\_dist/1000
        & \begin{aligned}
            & \text{Dynamic penalty based on distance } \\
            & \text{from goal} \\
        \end{aligned} \\
        \hline 
    r_{\text{braking}}=-1
        & \begin{aligned}
            & \text{Penalty if the car brakes when it’s not } \\
            & \text{ moving} \\
        \end{aligned} \\
        \hline 
    r_{\text{steer}}=-1
        & \begin{aligned}
            & \text{Penalty if  }\alpha_t \neq 0 \\
        \end{aligned} \\
  \end{cases}
\]

\end{definition}


One drawback of NavA2C \cite{gupta2022hybrid} is the consistent acceleration during the test phase when actions are determined deterministically. To address this issue, we mitigated $r_{goal\_reached}$ and introduced a penalty for overspeeding, denoted as $r_{over\_speeding}$. Drawing inspiration from HyLEAR \cite{gupta2022hybrid}, we incorporated $r_{near\_miss}$ into the reward function to penalize near misses, encouraging the car to maintain a safe distance from pedestrians. Additionally, $r_{hit}$ penalizes crashes or collisions with pedestrians, promoting the safety of the autonomous car. The term $r_{obstacle}$ is designed to encourage the autonomous car to avoid collisions with other obstacles. Furthermore, $r_{steer}$ promotes comfortable and smoother navigation by penalizing the agent for choosing actions that necessitate excessive steering. Finally, $r_{not\_goal}$ incentivizes reaching the goal more expeditiously.


\section{Methods}\label{sec:Methods}

In this section, we introduce Nav-Q, the first quantum-supported deep reinforcement learning architecture that combines classical and quantum neural networks to effectively train a reinforcement learning agent for safe navigation of self-driving cars. Nav-Q is primarily based on an actor-critic model, wherein the critic component is realized through a parameterized quantum circuit and a fully connected classical layer. This design enables us to harness the capabilities of a quantum algorithm exclusively during the training phase, eliminating the need for a QPU in the vehicle during testing.

Additionally, we introduce a novel quantum embedding technique that extends the Data Reuploading strategy to efficiently encode high-dimensional vectors using an arbitrary number of available qubits.

\subsection{Nav-Q Architecture}

Nav-Q is inspired by  NavA2C (Section \ref{sec:NavA2C}), which comprises two major components: an \textit{Anytime Weighted Hybrid A* Path Planner} \cite{hyleap}, determining the steering angle for the car, and DRL components—a hybrid actor-critic model responsible for determining speed actions (accelerate, maintain, decelerate). 
The objective is to reach the destination in the shortest time possible while avoiding collisions with other agents that may be present in the scene (such as other cars, pedestrians, etc.).

To be able to generalize any possible scenario that the car might encounter, the DRL agent needs to be trained for thousands of episodes, which implies running the training procedure for several days even when considering a limited set of scenarios. Furthermore, DRL algorithms are very unstable with respect to weight initialization, which means that it is necessary to run the training procedure several times with different initial parameters to achieve reliable performance in terms of driving policies.

The idea of Nav-Q is to introduce a quantum component into the DRL agent of NavA2C to obtain a quantum-supported architecture that can improve classical performance in terms of trainability and/or stability while having (at least) the same performance in terms of learned policy (time-to-goal, number of crashes, etc.). Also, the quantum component has not to be included at the time of testing, since this would require placing a QPU in the car, and invalidate the practical use of the proposed methods. 

\begin{figure}
    \centering
    \includegraphics[width=1\linewidth]{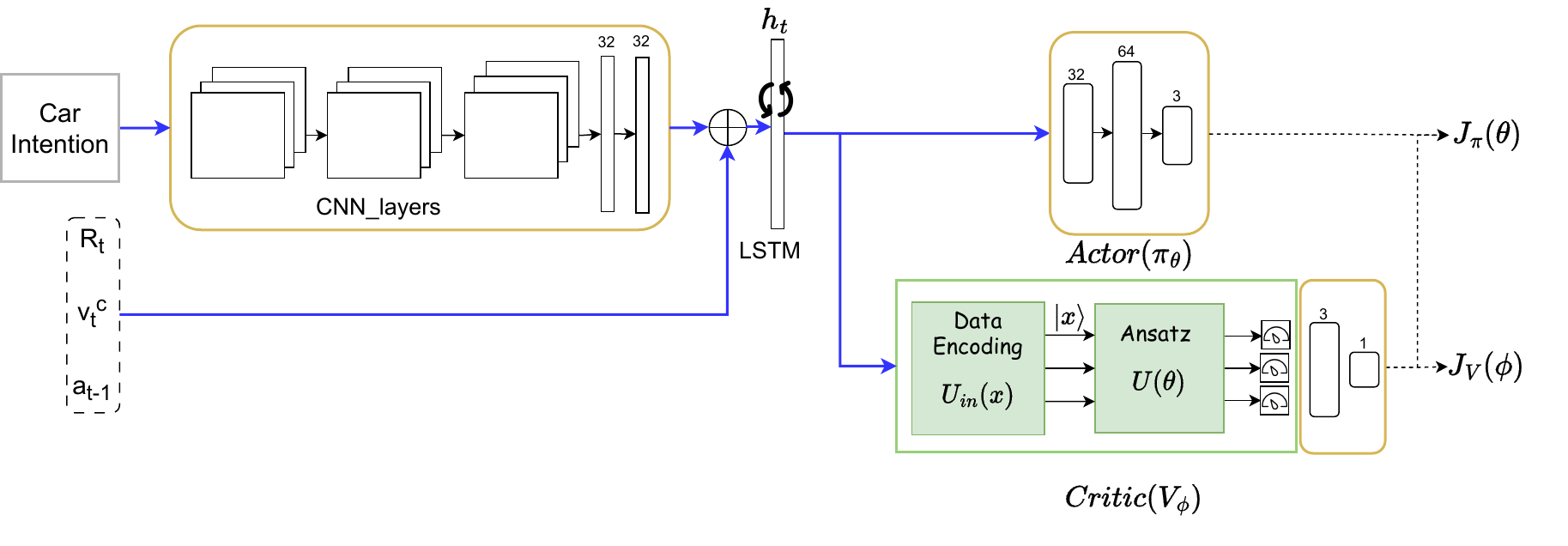}
    \caption[Quantum-supported DRL agent of Nav-Q]{Quantum-supported Deep Reinforcement Learning agent in Nav-Q. In contrast to the complete architecture of NavA2C illustrated in Figure \ref{fig:NavA2C architecture}, this depiction focuses solely on the DRL agent. The primary distinction from the classical model lies in the DRL agent's critic, which is replaced by a hybrid quantum-classical algorithm. Its design is essential for training and evaluating the loss function.}
    \label{fig:Nav-Q}
\end{figure}

To fulfill these requirements, Nav-Q introduces a parametrized quantum circuit to implement the critic within the actor-critic framework. This strategic choice allows us to exclusively employ the quantum algorithm during the training phase, where the critic evaluates the policy selected by the actor. In contrast, during the testing phase, only the actor is utilized, as it produces the action.
Therefore, the training process, recognized as the most challenging aspect of RL, is where the introduction of quantum components is expected to have the maximum impact. 

The quantum-supported DRL agent of Nav-Q is visually represented in Figure \ref{fig:Nav-Q}.

As for NavA2C, the Nav-Q agent  
comprises four main components: Feature Encoder $E_\psi$, Actor $\pi_\theta$, LSTM, and Critic $V_\phi$. 
$E_\psi$ is responsible for learning a low-dimensional representation of car intention $I$, using a long sequence of CNN layers. The output of the feature encoder, $E_\psi(I)$ is concatenated with the reward $R_t \in \mathbb{R}$, the car's velocity $v_t \in \mathbb{R}^2$, and the car's speed action in the previous step $acc_{t-1} \in \mathbb{R}$ and passed as input to the LSTM (See Figure \ref{fig:Nav-Q}). The LSTM, stores in its hidden state a more compressed representation of past observations and current input. It encodes information about the previous states that the agent has encountered, helping to infer the current state even when the observations are noisy and incomplete. The LSTM's hidden state $h_{t}$ is then passed to the actor and critic networks to determine $\pi_\theta(a_t | h_t, o_t)$ and $V_\phi(h_t, o_t)$ respectively. Recall that the Value function $V_\phi(h_t, o_t)$ implemented using the critic network gives the expected return if one starts in state $s$ and always acts according to policy $\pi_\theta$.
The Actor ($\pi_\theta$) in Nav-Q is implemented through a dense neural network and is the same as to NavA2C. It takes input $h_t$ from the LSTM layer and produces a $3$-dim vector with associated probabilities for speed actions (accelerate, maintain, and decelerate) during training. 

The implementation of the critic constitutes a key distinction compared to NavA2C and, more broadly, to all classical actor-critic approaches.


\subsection{Hybrid Quantum-Classical Critic}
The critic $V_\phi(h_t, o_t)$ in Nav-Q is realized through a hybrid quantum-classical algorithm made up of two components. The first one is a PQC which processes  $h_t$ - the hidden state of the LSTM - through unitary operations to generate a parametrized quantum embedding in the Hilbert space.  The quantum embedding is then converted into an $n$-dimensional classical vector (where $n$ is the number of qubits of the quantum circuit) using the expectation measurement of Pauli-$Z$ observables for each single qubit. 
At this point, there is a mismatch between the output of PQC ($\mathbb{R}^n$) and the required dimension of the Value function $(V_\phi(h_t, o_t) \in \mathbb{R}$). Thus, to map the $n$-dim vector resulting from the PQC to a single value of $V_\phi$, the second part of the critic is implemented, which is essentially a fully connected (FC) neural network layer made up of $n$ input and one output nodes (no hidden layers).

Despite the reduction in problem dimensionality given by the combination of the feature encoder and the LSTM, the complexity of solving real-world issues, such as the CFN, poses a challenge in terms of data encoding within a parametrized quantum circuit.
To tackle this challenge, we propose the \textit{Qubit Independent Data Encoding \& Processing} (QIDEP) strategy, which enables the encoding of a high-dimensional vector using an arbitrary number of qubits.

Note that the utilization of $\theta$ parameters for the actor and $\phi$ parameters for the critic aligns with established RL literature on actor-critic approaches. However, starting from the next section, we employ the parameters $\theta$ to denote the trainable parameters of the quantum circuit implementing the critic to be consistent with quantum computing literature.

\subsection{Qubit Independent Data Encoding \& Processing}

Data reuploading \cite{P_rez_Salinas_2020} is an efficient strategy for quantum embedding which allows the approximation of any continuous function using a single-qubit circuit. We extend this idea to propose a novel strategy, named Qubit Independent Data Encoding and Processing (QIDEP), where a multi-qubit system allows encoding a high-dimensional vector using a flexible number of qubits. A formal definition of this strategy is given in Definition \ref{def: QIDEP}. 
\begin{figure}
    \centering
    \includegraphics[scale=.95]{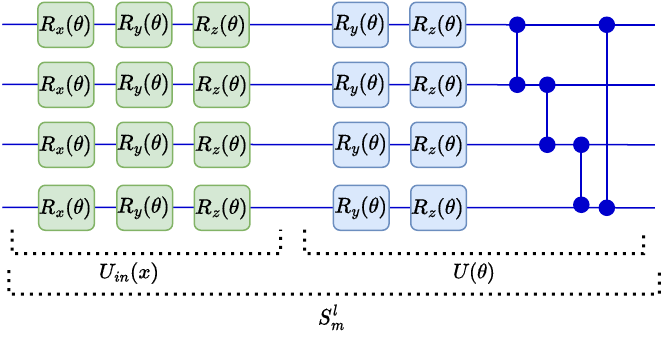}
    \caption{Graphical representation of the sublayer concept in QIDEP, comprising $U_{in}(x)$ and $U(\theta)$. The quantum gates of $U_{in}(x)$ for data encoding are illustrated in green, while the ansatz $U(\theta)$ with trainable parameters is depicted in blue. In this example, the $U(\theta)$ structure adopts a hardware-efficient architecture, consisting of parameterized $R_y$ and $R_z$ gates, followed by a daisy chain of CZ gates. On the other hand, the quantum gates of $U_{in}$ align with those proposed in the classical Data Reuploading technique \cite{P_rez_Salinas_2020} consisting of a sequence of the three Pauli rotation gates, whose parameters are non-trainable and depend on the input feature $x$.
    }
    \label{fig:sublayer}
\end{figure}
\vspace{.5em}

\begin{definition} [Qubit Independent Data Encoding \& Processing] \label{def: QIDEP}

Let $x \in \mathbb{R}^p$ be the input classical data, $U(\theta)$ be a parameterized quantum circuit with trainable parameters $\theta$, $U_{in}(x)$ be the unitary for data encoding based on Angle Encoding, $n$ be the desired number of qubits and $L \in \mathbb{N}$ be the desired number of layers in the quantum circuit. 
In QIDEP, each layer \(U_l\) can be written as a composition of sublayers \(S_m^l\) 
\begin{align}
U_l = \prod^{m=1}_{m=k} S_m^l 
\end{align}
where $S_m^l$ is the $m$-th sublayer of the $l$-th layer defined as $S_m^l = U(\theta_m^l)U_{in}(x^{m})$, with $x^m = \{x_{3n(m-1)+1}, \dots , x_{3nm}\}$ indicating a specific subset of elements in the input feature $x$. Note that the number of sublayers \(k = \lceil\frac{p}{3n}\rceil\) is fixed and given the input feature and the number of qubits, the vector must be padded with a zero vector of length \(3nk-p\).

Importantly, the architecture allows to freely choose the number of layers, as well as the number of trainable parameters within each sublayer, depending on the task at hand. Thus, a $L$-layered, $n$-qubit QIDEP quantum circuit that has to process a $p$-dimensional vector can be written as: 
\begin{align}
    U(\theta, x) = \prod^{l=1}_{L} \prod^{m=1}_{m=k} U\left(\theta_m^l\right)U_{in}\left(x^m \right)  
\end{align}

\end{definition}

\vspace{.5em}

\begin{figure}
    \centering
    \includegraphics[width=1.0\linewidth]{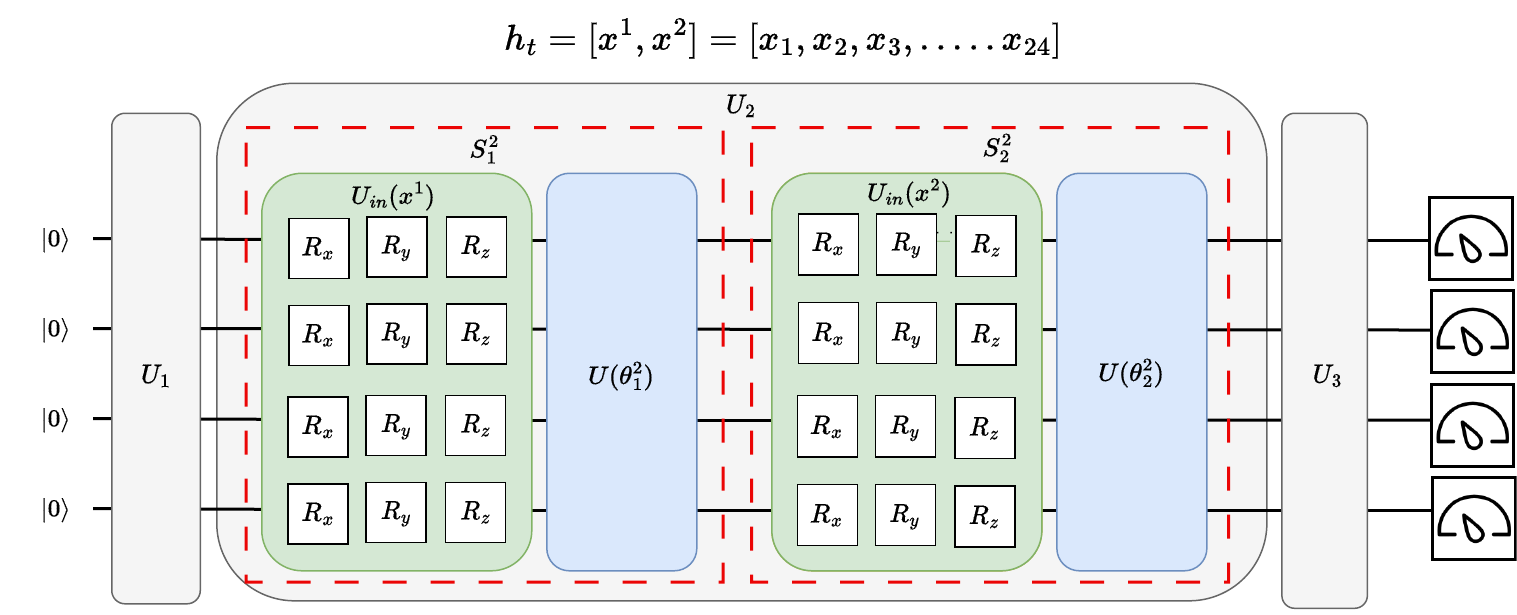}
    \caption{Example of a quantum circuit implementing the QIDEP strategy to encode a $24$-dimensional vector using $4$ qubits and $3$ layers. The quantum circuit is segmented into layers, each containing identical quantum gates but varying parameters within each ansatz. Each layer comprises sublayers $S^l_m$, which, in turn, includes a data encoding gate, $U_{in}(x)$, and the ansatz, $U(\theta)$. The grey boxes represent the layers, the red dotted rectangles symbolize sublayers, the green boxes denote the data encoding module $U_{in}$, and the blue boxes represent the ansatz $U(\theta)$. Notice that the additional classical fully connected layer connected to the measurements of the $4$ qubits is not part of the specific QIDEP design but rather a component of the Nav-Q architecture.
    }
    \label{fig:example_data_reuploading}
\end{figure}

An example of QIDEP strategy is shown in Figure \ref{fig:example_data_reuploading}.
In Data Reuploading \cite{P_rez_Salinas_2020}, each layer can be seen as a single unit of $U_{in}$ and $U(\theta)$, and the number of layers in the circuit is considered a hyperparameter (as in classical neural networks). In QIDEP, we introduce the concept of sublayer $S$. Specifically, each layer $l$ consists of $k$ sublayers ${S^l_m}$, each comprising unitaries $U_{in}(x)$ and $U(\theta)$. Here, $U_{in}(x)$ is employed to encode a contiguous sub-sequence of features of $x$, while $U(\theta)$ represents the parametrized ansatz.
Figure \ref{fig:sublayer} shows an example of a sublayer considering a 4-qubit circuit and a hardware efficient \cite{Kandala_2017} parametrized ansatz.

Specifically for the Nav-Q architecture, the dimensionality reduction task is implemented through a Fully Connected (FC) layer \cite{lan2021variational,lockwood2020reinforcement}. Incorporating an FC layer for dimensionality reduction helps scale the output of the PQC using trainable weights, enabling the critic to achieve a Value function output that aligns more closely with the range of the true Value function. Additionally, a trainable output range can enhance convergence and reduce sensitivity to the choice of hyperparameters and random initialization of model parameters \cite{Skolik_2022}.

As a result, the second part of the critic is implemented using an $n\times 1$ FC neural network layer. This layer processes the $n$-dimensional vector, generating a single number that represents the Value for the observation $o_t$. An $n\times 1$ FC layer involves a total of $n+1$ trainable parameters—$n$ weights and $1$ bias.

\section{Performance Evaluation}\label{sec:Experiments}

In this section, we assess Nav-Q using the CARLA-CTS benchmark, a widely recognized virtual environment for evaluating the performance of  DRL models in the context of self-driving cars. Additionally, we compare the results of Nav-Q with those of its classical baseline, NavA2C, which relies solely on conventional computational resources. Finally, we train Nav-Q on a simulated noisy quantum system with the parameter-shift rule to estimate the expected performance on real quantum hardware.

\subsection{Experimental Settings}

\paragraph{Environment}

\begin{figure}
    \centering
    \includegraphics[width=1\linewidth]{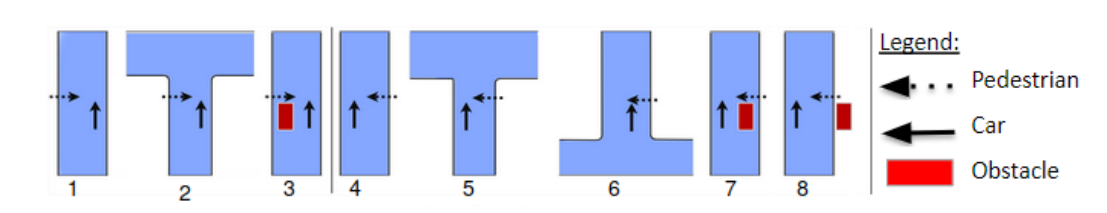}
    \caption[Accident scenarios on the basis of GIDAS study]{Accident scenarios on the basis of GIDAS \cite{GIDAS} study. The autonomous vehicle is represented by the solid lines, stationary objects that block its view are represented by the red squares, and pedestrian movement is represented by the dotted lines. Scenario id is represented by the numbers at the bottom of each scenario.}
    \label{fig:gidas}
\end{figure}

For comparative experimental evaluation of CFN methods of self-driving cars in critical traffic scenarios, we use the virtual CARLA-CTS benchmark \cite{gupta2022hybrid}, an extension of OpenDS-CTS \cite{hyleap} for the driving simulator CARLA \cite{dosovitskiy2017carla}. This benchmark contains about thirty-thousand scenes of twelve scenarios mainly based on the GIDAS \cite{GIDAS} study of road accidents with pedestrians in Germany and simulated in CARLA on a test drive of about 100 meters. In Figure \ref{fig:gidas}, the blue boxes with solid arrows denote autonomous car movements, the red boxes with solid arrows denote static or dynamic obstacles such as parking or incoming cars, and the dotted arrows denote the pedestrian movement. 


 Since Nav-Q is trained using quantum simulations, the training time is considerably high, thus to to reduce the complexity of the tested environment, we adopt as training scenarios  1, 3, 4, 5, 6, and 8 and test our model on all scenarios from 1-8. A scene is represented by a 6-tuple $(\text{scenario}, p^c_{\text{start}}, p^c_{goal}, \theta^c_\text{start}, P, C)$, where $\text{scenario} \in \text{Scenarios}$ correspond to the scenario being simulated in the scene, $p^c_{\text{start}} \in \mathbb{R}^2$ denotes the starting position of the car,  $\theta^c_\text{start} \in \mathbb{R}^2$ represents the starting orientation of the car,  $p^c_{goal} \in \mathbb{R}^2$ indicates the goal position of the car, $P$ is the set of pedestrians in the environment and $C$ is the set of cars in the environment apart from the autonomous car $c$. Each scene is essentially an instantiation of a particular scenario with a specific configuration for pedestrian spawn point and pedestrian speed. For training, we consider pedestrian speed in the range of $0.6-2.0$ m/s with a step size of $0.1$ and pedestrian distance from $0-40$ m with a step size of $1$. For testing, we consider pedestrian speed in the $0.25-2.85$ m/s range and pedestrian distance in the $4.75-49.25$ m range. As a result, there were $3690$ scenes in the training set and $9936$ scenes in the test set.  

\paragraph{Classical Baseline} \label{sec: NavA2C}

In this study, we evaluate the performance of Nav-Q in comparison to its classical counterpart, NavA2C, which draws inspiration from A2C-CADRL-p \cite{gupta2022hybrid} (Section \ref{sec:NavA2C}). Notably, we introduce several enhancements that aim to improve its performance compared to earlier iterations.

One key modification involves the incorporation of layer normalization \cite{ba2016layer} into the neural network architecture of the Feature Encoder ($E_\psi$), Actor ($\pi_\theta$), and Critic ($V_\phi$). This addition was absent in the original A2C-CADRL-p model \cite{gupta2022hybrid}. The inclusion of layer normalization is intended to promote more stable and expedited training. 
Furthermore, in the initial version of A2C-CADRL-p, the LSTM produced a 256-dimensional vector ($h_t \in \mathbb{R}^{256}$). Given the challenges of encoding such high-dimensional vectors into a quantum state, we made adjustments to $E_\psi$ and LSTM to produce a 32-dimensional vector instead. Additionally, we fine-tuned the reward function to facilitate more consistent and stable learning (See Section \ref{sec:NavA2C}).

It is worth noting that, in the realm of CFN for self-driving cars, our study is pioneering in its investigation of hybrid quantum-classical reinforcement learning algorithms. As a result, there are no quantum baselines to compare with, as we are the first to delve into this domain.

\paragraph{Implementation}

To implement the parametrized quantum circuit in the critic of Nav-Q, we conducted experiments using 4 and 6 qubits, varying the number of layers from 1 to 3. To select the best quantum model, we trained different configurations for up to 10,000 episodes and chose the configuration that yielded the best results in terms of policy improvement. Once the best quantum model was identified, we trained Nav-Q and NavA2C for 10,000 episodes to assess the performance in terms of training policy of both approaches. To evaluate convergence rate and training stability we run six different weight initializations of the overall architecture, training for up to $5,000$ episodes, and analyzing the average training behavior. 

For both NavA2C and Nav-Q, gradients were computed after each episode, and the network weights were updated using the Adam optimizer with a learning rate of $0.0005$. Each episode had a maximum duration of $500$ time steps and ended prematurely if the car reached its goal or collided with any obstacle. Given the challenge of encoding high-dimensional classical data into a quantum state, we set the output of the LSTM layer for both Nav-Q and NavA2C to produce $h_t \in \mathbb{R}^{32}$, which was initially a $256$-dimensional vector in A2C-CADRL-p \cite{gupta2022hybrid} (the inspiration for NavA2C).

In the case of Nav-Q, we utilized the `default.qubit' device of PennyLane for quantum simulation and employed backpropagation with gradient descent to update the weights of the entire model. It's important to note that using backpropagation for training quantum circuits would not be feasible on quantum hardware \cite{Mitarai_2018,Schuld_2019}. To estimate the performance as if we had a real quantum device, we also trained Nav-Q using the parameter-shift rule \cite{Mitarai_2018}, an alternative method to compute partial derivatives of quantum expectation values with respect to gate parameters on quantum hardware. Nevertheless, gradient calculation through the parameter-shift rule is computationally expensive. To address this, we reduced the dimension of the LSTM output from $32$ to $6$ (i.e., $h_t \in \mathbb{R}^{32}$) and trained a circuit with only $2$ qubits and $1$ layer to reduce circuit depth. We also simplified the learning task by training on only the first scenario of the CARLA-CTS 1.0 benchmark for $2,500$ episodes.

Finally, we also trained and tested the simplified version of Nav-Q under two simulated noise models, Quantum Gate Error \cite{pyquil-noise} and Depolarising Error \cite{king2002capacity}.  To simulate Quantum Gate Error, we add uniform random noise to the parameters of rotation gates such that $\theta \leftarrow \theta + 0.01\times\theta\times \delta$, where $\delta\sim \text{Uniform}(0, 1)$. This kind of error arises due to imperfection in the realization of rotation gates on quantum hardware. Depolarising Error represents a random error in between each operation on a qubit. These errors are injected in the form of bit-flips (x error), phase-flips (z error), or both at the same time (y error). We use the in-built library of Pennylane \cite{bergholm2022pennylane} to simulate this error.

\paragraph{Metrics}

To assess the performance Nav-Q with the classical baselines, we define three categories of metrics:

\begin{enumerate}
    \item \textbf{Trainability:} 
This aspect refers to the capacity of an RL model to be effectively trained to perform a specific task or acquire a particular behavior within the environment. A conventional method for evaluating trainability involves analyzing the \textit{Return vs. Episodes} curve, illustrating the smoothed cumulative reward (i.e., return) as the number of episodes increases. Specifically, we assess the mean and standard deviation of the corresponding Area Under the Curve (AUC) \cite{chen2020variational, Skolik_2022} to comparatively evaluate the overall convergence rate and stability across various initializations of the overall architecture. 

    \item \textbf{Driving Policy}:
    The performance of each method in terms of driving policy is measured using the following metrics: (a) the overall safety index (SI) defined as total number of scenarios in which the method is below given percentages of crashes and near-misses; (b) the crash and near-miss rates (\%), and time to goal (TTG) in seconds \cite{gupta2022hybrid}.

    \item \textbf{Learnability}: To methodologically assess the ability of both quantum and classical models to learn complex and diverse functions beyond the immediate problem statement, we employ the calculation of the effective dimension \cite{berezniuk2001scale,Abbas_2021}. This complexity measure, rooted in information geometry, estimates the size of the space containing all possible functions for a particular model class, utilizing the Fisher information matrix (FIM) as the metric. Moreover, the spectral analysis of the FIM enables the examination of the curvature of loss surfaces, providing insights into the nature of loss surfaces for both classical and quantum models \cite{Abbas_2021}. Indeed, the shape of the optimization landscape plays a crucial role in understanding whether a specific model can effectively find a suitable solution to the problem.

 \end{enumerate}

\subsection{Results}

\paragraph{Trainability}

\begin{figure}[h]
\centering
\begin{tabular}{cc}
\hspace{-3em}
\includegraphics[scale=0.45]{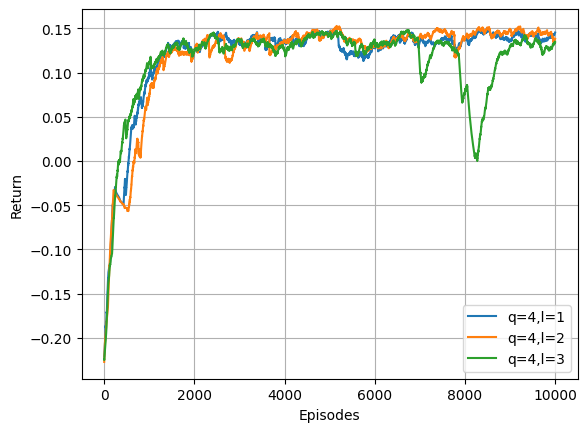}
&
\includegraphics[scale=0.45]{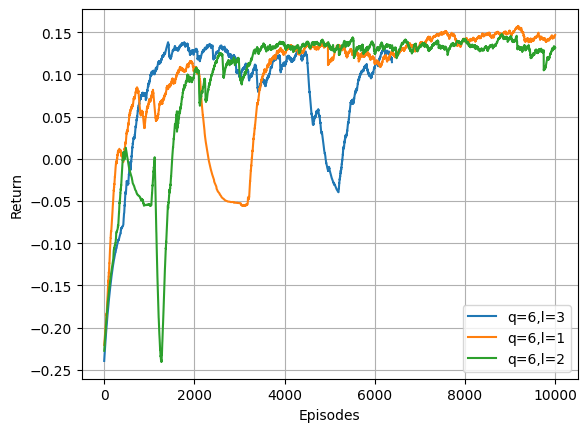}
\end{tabular}
\caption{\textit{Return vs. Episodes} smoothed curves of Nav-Q with a 4-qubit (\textbf{left}) and 6-qubit (\textbf{right}) parametrized quantum circuit and number of layers ranging from $1$ to $3$.}
\label{Fig:best_Nav-Q_Selection}
\end{figure}

To choose the optimal architecture for Nav-Q and compare it with the classical baseline, we ran various configurations of the PQC for the critic, varying the number of qubits and layers and assessed the convergence rate and stability in terms of the \textit{Return vs. Episode} curve. Results are shown in Figure \ref{Fig:best_Nav-Q_Selection}.

\begin{figure}[ht]
    \centering
    \includegraphics[width=0.68\linewidth]{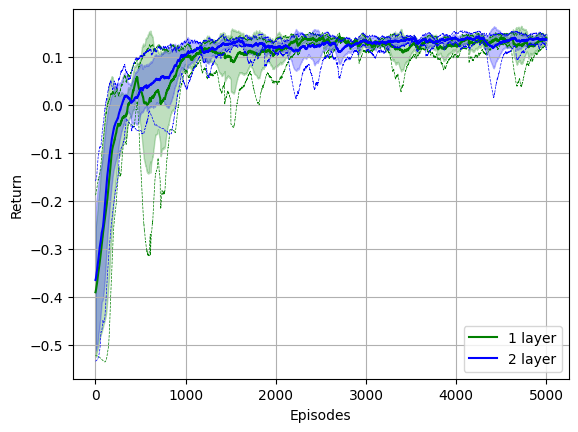}
    \caption{Comparison of \textit{Return vs. Episodes} graph of two different 4-qubit Nav-Q architectures with 1 and 2 layers respectively, executed for six different parameters initializations.}
    \label{fig:l1vsl2}
\end{figure}

We observe that, in all cases, the model converges after 10,000 episodes, yielding comparable return, despite having varying numbers of parameters. This aligns with a well-known observation in the classical domain: an increase in the model's complexity does not necessarily guarantee a performance improvement. However, when using a 6-qubit critic, training performance becomes unstable, with a sudden drop in cumulative reward (\textit{return}) requiring hundreds of episodes to recover an increasing trend. This instability is not observed with a 4-qubit circuit using one or two layers. The behavior can be attributed to a high number of parameters leading to optimization issues arising from a complex landscape, potentially affected by barren plateaus. 
Table \ref{tab:best_model} presents a comparison of training times and the number of parameters for various configurations of Nav-Q. The training times show an increase with the number of qubits and layers, posing a significant challenge and extending to several days for each model. Nevertheless, when addressing real-world problems, such as the CFN benchmarked on standard classical data, testing a quantum-supported model remains feasible, albeit challenging.
\begin{table}[ht]
\begin{tabular}{ccccc}
\toprule
\textbf{Method} &
  \textbf{\#qubits} &
  \textbf{\#layers} &
  \textbf{\begin{tabular}[c]{@{}l@{}}\#parameters ($d$)\end{tabular}} &
  \textbf{\begin{tabular}[c]{@{}c@{}}Training Time \\ (days)\end{tabular}} \\ \midrule
\multirow{6}{*}{Nav-Q} & \multirow{3}{*}{4}  & 1  & 29 & 6.8 \\ 
                       &   & 2  & 53    & 10.87 \\
                       &   & 3  & 77    & 14.64  \\
                       & \multirow{3}{*}{6}  & 1  & 31   & 8.14  \\
                       &   & 2  & 55   & 11.74  \\
                       &   & 3  & 79   & 18.46  \\
\botrule
\end{tabular}
\caption{Training times and number of parameters for different critic architectures of Nav-Q.}
\label{tab:best_model}
\end{table}
Considering the comparable results of the 4-qubit architecture, particularly when focusing on configurations with only one or two layers, the key consideration becomes the stability of the model in choosing the optimal architecture.
In this respect, we conducted an analysis using five different weight initializations. The results, encompassing mean, standard deviation, and range, are illustrated in Figure \ref{fig:l1vsl2}. This choice is motivated by the imperative to reduce training times, as emphasized in Table \ref{tab:best_model}, where each experiment demands several days for a single training session to be completed.

Despite the similarity in average performance in terms of return for both architectures, we observe that the 1-layer model exhibits less stability compared to its 2-layer counterpart. Given that stability, along with average reward, is a crucial factor in analyzing reinforcement learning performance, we choose the critic with a parameterized quantum circuit using 4 qubits and 2 layers as the best model.
\begin{figure}[ht]
    \centering
    \includegraphics[width=0.68\linewidth]{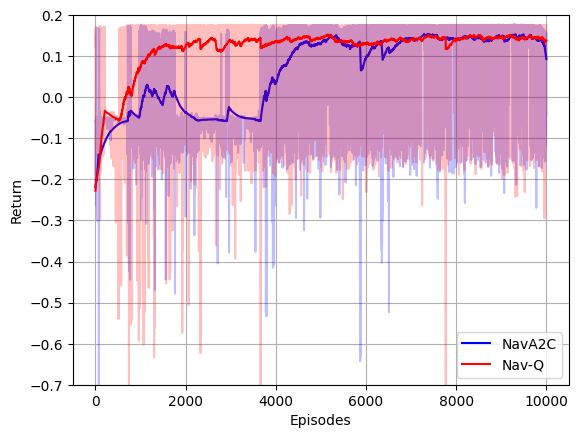}
    \caption[Comparison of the smoothed \textit{Return vs. Episodes} curves of NavA2C and Nav-Q with 4 qubits and 2 layers for a single run]{Comparison of \textit{Return vs. Episodes} graph of NavA2C and Nav-Q with 4 qubits and 2 layers. }
    \label{fig:BestQuantumvsClassical}
\end{figure}

The next step is the comparison of the best configuration of Nav-Q with its classical counterpart NavA2C. In particular, we trained NavA2C for $10,000$ episodes and compared it with Nav-Q. Results are shown in Figure \ref{fig:BestQuantumvsClassical}.

\begin{figure}[ht]
    \centering
    \includegraphics[width=0.7\linewidth]{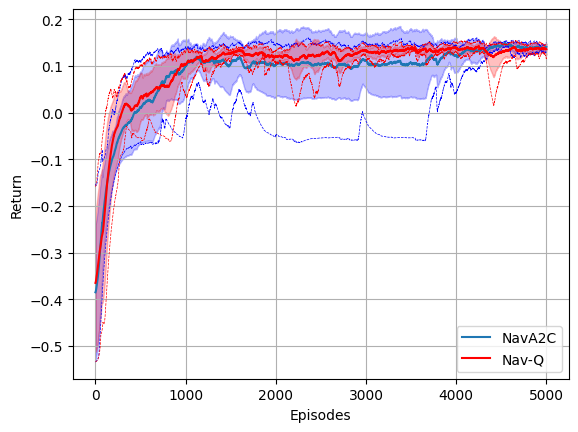}
    \caption{Comparison of the smoothed \textit{Reward vs. Episodes} curves for  Nav-Q with $4$ qubits and $2$ layers, and NavA2C averaged over multiple runs. The red-colored plots in the graph correspond to Nav-Q while the blue colored plots refer to NavA2C.  The thick solid line represents the mean of $6$ runs, each with a different weight initialization of the overall architecture. The colored shaded area represents the standard deviation and the thin dotted line represents the maximum and minimum return values. }
    \label{fig: NavQvsNavA2C}
\end{figure}

Nav-Q exhibits a superior convergence rate in the initial $4,000$ episodes, indicating that a DRL agent with an enhanced quantum critic demonstrates faster trainability compared to the classical counterpart. However, over an extended duration, the two models attain similar performance.
To better assess the convergence rate and the training stability, we conducted six training sessions each with distinct initialization weights. This experiment aims to evaluate two critical aspects: the average convergence rate (performance improvement as the number of episodes increases) and stability (sensitivity of the two models to different weight initializations of the neural networks, both classical and quantum).
Results are shown in Figure \ref{fig: NavQvsNavA2C}.
The trend of the curves remains consistent for both quantum and classical models, suggesting similar overall performance in terms of driving policy, as evidenced by the resemblance in average rewards between the two approaches. However, the Nav-Q return is mostly higher than NavA2C, albeit to a lesser extent than in the previous case.

To quantitatively evaluate convergence and stability, we calculate the average and standard deviation of the AUC of the \textit{Return vs. Episode} curve across the six runs. A higher average AUC implies better convergence, while a lower standard deviation indicates enhanced stability.

Nav-Q reports a higher average AUC ($2440.6$) than NavA2C ($2362.95$). Upon examining the standard deviation, Nav-Q demonstrates lower variability $(758.78)$ compared to NavA2C $(902.73)$, which implies better stability of the quantum model with respect to its classical counterpart. 

\paragraph{Driving Policy}

Starting with the two models trained for $10,000$ episodes (Figure \ref{fig:BestQuantumvsClassical}), we evaluated the performance in terms of driving policy. Results are shown in Table \ref{tab: driving policy}.
\begin{table}[ht] 
\begin{tabular}{cccccccc}
\toprule
\textbf{Method}       & \textbf{\begin{tabular}[c]{@{}l@{}}\#para-\\ meters\end{tabular}} & \textbf{\begin{tabular}[c]{@{}l@{}}Time\\ to\\ goal\\ (s)\end{tabular}} & \textbf{\begin{tabular}[c]{@{}l@{}}crashes \\ (\%)\end{tabular}} & \textbf{\begin{tabular}[c]{@{}l@{}}near-\\ misses\\ (\%)\end{tabular}} & \textbf{\begin{tabular}[c]{@{}l@{}}Safety\\ Index\\ (SI)\end{tabular}} & \textbf{\begin{tabular}[c]{@{}l@{}}Mean\\ Return\end{tabular}} & \textbf{\begin{tabular}[c]{@{}c@{}}Training Time \\ (days)\end{tabular}}\\ 
\midrule
Nav-Q  & 53   & \textbf{11.74}   & 13.23  & 23.62  & 3 & 0.10 & 10.87\\ 
NavA2C & 2305 & 11.77 & \textbf{12.47}  & \textbf{22.39}  & 3  & \textbf{0.12} & \textbf{2.7} \\ \botrule
\end{tabular}
\caption{Comparison of driving policy of NavA2C and Nav-Q with 4 qubits and 2 layers after 10,000 episodes of training.}
\label{tab: driving policy}
\end{table}

Both approaches seemed to have learned a similar policy. Nav-Q has a slightly lesser mean time to goal compared to NavA2C, which implies that the model trained using a quantum critic reaches the goal faster compared to the fully classical approach. When looking at collisions (crashes and near-misses), the classical approach slightly outperforms Nav-Q, providing a lower percentage for crashes and near-misses. However, the SI -defined in the range $[1, 8]$ that quantifies the total number of scenarios in which the agent exhibits the number of crashes and near misses below $20 \%$- is the same. In general, the differences in terms of driving policy seem to be not very significant when comparing Nav-Q with NavA2C. This implies that the advantages provided by Nav-Q in terms of convergence rate and stability come without affecting the performance when considering the behavior of the car during the test phase.

Finally, the last column of Table \ref{tab: driving policy} reports the number of days required to train the two models. As observed, while the quantum-supported architecture reduces complexity in terms of the number of parameters, the training times are notably longer. This suggests that the current optimization strategy, which includes the forward pass—where it is necessary to simulate the full quantum circuit after each update—and standard backpropagation, which is usually well-suited for classical neural networks, might require custom amendments to perform optimally when dealing with quantum-supported solutions on real-world data.

\paragraph{Learnability}

The normalized effective dimension (ED) \cite{Abbas_2021}, denoted by $\overline{d}{\gamma, n}= \frac{d{\gamma, n}}{d}$ provides the proportion of active parameters relative to the total number of parameters in the model. A high normalized effective dimension suggests that the model is utilizing a substantial portion of the model space, with many parameters actively involved in fitting the data. Conversely, a low effective dimension indicates that the model is using only a limited part of the model space, with a smaller proportion of parameters actively contributing to the model's performance. Results comparing Nav-Q and NavA2C in terms of ED are shown in Table \ref{tab:ED}.
\begin{table}[ht]
\begin{tabular}{cccccc}
\toprule
\textbf{Method} &
  \textbf{\#qubits} &
  \textbf{\#layers} &
  \textbf{\begin{tabular}[c]{@{}l@{}}\#parameters ($d$)\end{tabular}} &
  \textbf{ED ($d_{\gamma, n}$)} &
  \textbf{\begin{tabular}[c]{@{}l@{}}Normalized ED\\ $\overline{d}_{\gamma, n}= \frac{d_{\gamma, n}}{d}$\end{tabular}} \\ \midrule
\multirow{6}{*}{Nav-Q} & 4  & 1  & 29   & 8.14  & 0.28 \\
                       & 4  & 2  & 53   & 10.79 & 0.20 \\
                       & 4  & 3  & 77   & 13.30 & 0.17 \\
                       & 6  & 1  & 31   & 6.76  & 0.22 \\
                       & 6  & 2  & 55   & 9.11  & 0.17 \\
                       & 6  & 3  & 79   & 12.97 & 0.16 \\
NavA2C                 & NA & NA & 2305 & 56.65 & 0.02 \\
\botrule
\end{tabular}
\caption{Comparison of all the configurations tested for Nav-Q and NavA2C in terms of the number of parameters,  and the effective dimension. }
\label{tab:ED}
\end{table}
 We observed that $\overline{d}_{\gamma, n}$ for all circuits of Nav-Q is higher than NavA2C. This implies that the quantum critic has a greater capacity to model complex functions compared to the classical critic. Understanding the capacity of a quantum model can be advantageous as it provides insight into the model's capability to learn patterns even before training, which is a resource-intensive task.

\begin{figure}[ht]
  \centering
  \small NavA2C \hspace{15em} \small Nav-Q \\
  \vspace{.1em}
  
  \begin{tabular}{@{}c@{}}
    \includegraphics[width=.48\linewidth,height=130pt]{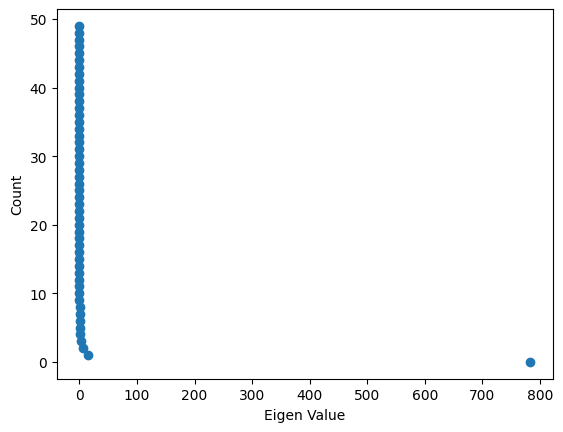} \\
  \end{tabular}
  \vspace{\floatsep}
  \begin{tabular}{@{}c@{}}
    \includegraphics[width=.48\linewidth,height=130pt]{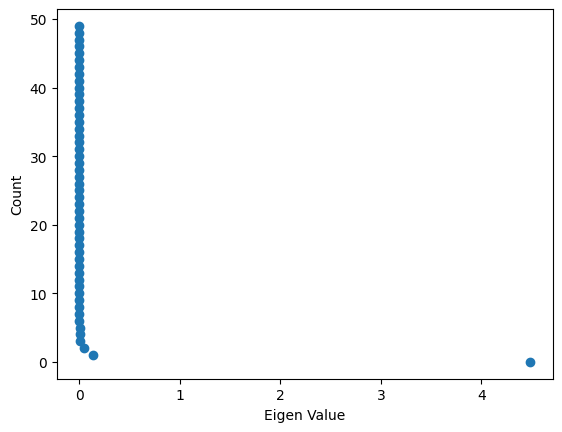} \\
  \end{tabular}
  \vspace{-1em}
  \caption{Eigenvalue distribution of Fisher Information Matrix for the critic of NavA2C (right) and Nav-Q (left).}
\label{Fig:EVD}
\end{figure}

In addition, we analyze the eigenspectrum of the FIM to gain insights on the shape of the optimization landscape. 
Figure \ref{Fig:EVD} shows the eigenvalue distribution of FIM obtained for the critics of Nav-Q and NavA2C.

We arrange the eigenvalues in descending order and plot the top $50$ eigenvalues for both cases. The subplots provide a closer examination of the distribution of eigenvalues that are close to zero. In the case of NavA2C, a few exceptionally high eigenvalues are observed, with the majority clustering near zero. This suggests that the loss curvature is steep along the eigenvector corresponding to the highest eigenvalue. Given that most eigenvalues are close to zero, we can infer that the loss surface is primarily flat, featuring minor ridges and valleys.

For Nav-Q, similar to NavA2C, a single prominent eigenvalue is noticeable, with the majority of the eigenvalues clustered close to zero. This phenomenon results in extremely small gradients, thereby slowing down optimization via gradient descent. This issue resembles the vanishing and exploding gradient problem encountered in classical neural networks.

This observation diverges from the findings in \cite{Abbas_2021}, where it is claimed that the eigenvalues in the case of PQCs should be more evenly spread. We speculate that the anomaly may be attributed to the benchmark design. However, we acknowledge that further investigation is required to fully understand this aspect.

\paragraph{Simulated Noisy Environment and Parameter-Shift Rule}

\begin{figure}[ht]
\centering
\begin{tabular}{cc}
\hspace{-2.3em}
\includegraphics[scale=0.45]{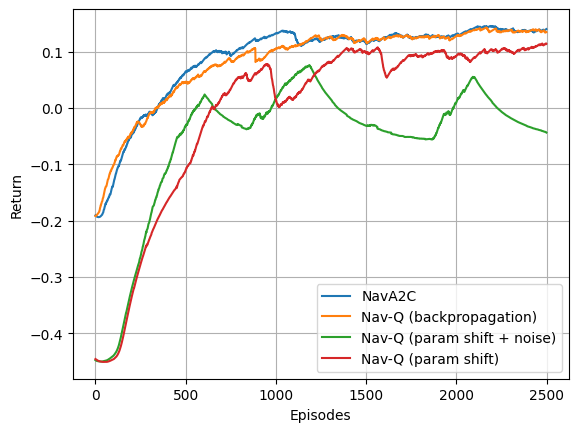}&
\includegraphics[scale=0.45]{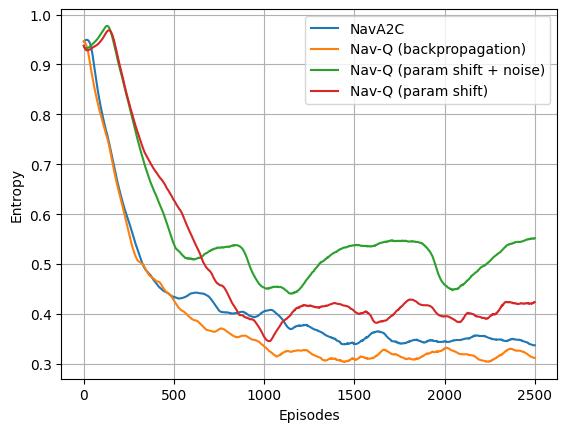}
\end{tabular}
\caption{Training performance in terms of \textit{Rewards vs Episodes} curve (left) and \textit{Entropy vs Episodes} curve (right) with respect to different optimizations settings.
In both graphs, we present the results for four different approaches: NavA2C represents the classical baseline, while two variants of Nav-Q are depicted depending on the optimization strategy adopted, classical backpropagation or parameter-shift rule. Additionally, one Nav-Q variant is trained with different noise models integrated in the circuit simulation, and the parameter-shift rule is employed for gradient calculation. }
\label{Fig:param_shift}
\end{figure}

All the previously described experiments utilized classical backpropagation for optimizing the parameters in Nav-Q. However, this approach is only feasible when using quantum simulation and would be impractical when dealing with real quantum devices. Furthermore, real near-term quantum devices are affected by noise which introduces errors and limits the coherence of quantum states, impacting the accuracy and reliability of quantum computations. To assess these two aspects, we conducted training procedures based on the parameter-shift rule and the introduction of noise. Figure \ref{Fig:param_shift} compares the different approaches using the \textit{Return vs. Episodes} and \textit{Entropy vs. Episodes} curves of NavA2C and Nav-Q.

Nav-Q trained using backpropagation and NavA2C exhibit similar performance, consistent with previous experiments. However, when training Nav-Q using the parameter-shift rule, the convergence rate is considerably slower than with backpropagation. In this regard, the entropy curve indicates that the parameter-shift rule is causing the problem of overshooting due to a large step size, as the entropy curve keeps oscillating around the value of $0.4$.

When introducing noise, we observe a deterioration in the stability of learning. The convergence rate of Nav-Q trained with the parameter-shift rule and introducing noise is almost the same as without noise in the initial part of the training. However, the problem of overshooting for the noisy model becomes even more pronounced, as evident from its entropy curve.
Therefore, we can conclude that, although the parameter-shift rule enables precise calculation of the gradient at each iteration on a methodological level, the performance remains below that of using backpropagation. This could be attributed to the need to set the magnitude of the gradient shift at each step in advance. Additionally, as expected, introducing quantum noise leads to a deterioration in performance in terms of both stability and convergence rate. Nevertheless, from the \textit{Entropy vs. Episodes} curve we observe that due to noise, the RL agent exhibits a higher exploratory nature since the value of entropy keeps oscillating even after a high number of episodes.

\section{Discussion and Conclusion}\label{sec:Conclusions}


In this work, we introduced Nav-Q, the first quantum-supported Deep Reinforcement Learning approach designed to address the challenge of Collision-Free Navigation for self-driving cars.
Nav-Q is built on an actor-critic algorithm to effectively learn an optimal policy for controlling the speed of self-driving cars. The critic is a hybrid quantum-classical algorithm constructed using an innovative technique named Qubit Independent Data Encoding and Processing (QIDEP), allowing the quantum embedding of high-dimensional classical data without being constrained by the number of qubits. Consequently, QIDEP provides the flexibility to adapt the number of qubits to suit specific quantum circuit requirements. Moreover, Nav-Q represents a valid alternative to existing classical algorithms, as its architecture design doesn't require a Quantum Processing Unit at test time, enhancing its usability for real-world applications.

To the best of our knowledge, Nav-Q is the first proposal of its kind, introducing a quantum component designed to enhance training performance without requiring a QPU during testing. This approach can be easily adapted to address typical problems utilizing the actor-critic model, particularly those struggling with trainability and stability during the training phase, and can derive benefits from quantum machine learning models.

We trained Nav-Q using the CARLA-CTS1.0 benchmark \cite{gupta2022hybrid} and systematically compared its performance with the classical baseline, NavA2C. In terms of trainability, the results demonstrated that, on average, Nav-Q exhibited a superior average convergence rate and better stability compared to NavA2C. Specifically, when running both approaches with different weight initializations, Nav-Q demonstrated a higher average AUC of the \textit{Return vs Episodes} curves than NavA2C, along with a lower standard deviation, implying better stability. We believe this is due to the higher number of parameters in NavA2C than in Nav-Q. Models with more parameters are generally more sensitive to trainable parameter initialization, as they have a larger space of possible configurations, making optimization more challenging. Considering the two models when trained for $10,000$ episodes, we evaluated their performance in terms of the driving policy of the car. In this regard, Nav-Q and NavA2C showed similar performance, with Nav-Q tending to achieve the final goal faster, while NavA2C was slightly better in terms of collision avoidance.  

In terms of learnability, Nav-Q exhibited superior performance when considering the effective dimension and showed similar behavior when examining the distribution of eigenvalues of the Fisher Information Matrix. In this context, the quantum model appears to be significantly impacted by the barren plateau problem, making the optimization procedure challenging in finding a suitable local minima.

Finally, we conducted smaller-scale training experiments with a $2$-qubit, $1$-layer Nav-Q using a suitable optimization procedure for real quantum hardware, known as the parameter-shift rule. We investigated the impact of noise using two distinct noise models: the Quantum Gate Error \cite{pyquil-noise} and Depolarising Error \cite{king2002capacity}. The comparison of the resulting \textit{Return vs. Episodes} curves clearly illustrates that the trainability of Nav-Q was adversely affected by the combination of the parameter-shift rule and noise, highlighting the challenges posed by these factors on near-term quantum hardware.

For this reason, a primary challenge to address in the near future is overcoming the issue of barren plateaus. Various approaches can be considered, including different parameter initialization techniques \cite{Grant_2019}, layerwise learning \cite{Skolik_2021}, and parameter correlations \cite{Holmes_2022}. Another interesting direction is to design a Quantum PPO architecture based on the principles of Nav-Q. In this case, it will be possible to leverage the efficiency of PPO in terms of more stable learning and faster convergence rate, while facilitating the training procedure through quantum computation.


Furthermore, a promising avenue involves extending the application of the Nav-Q architecture to diverse problems beyond the CFN of self-driving cars. The Nav-Q algorithm can be applied to problems requiring an RL agent challenging to train using classical neural networks alone. As observed, even a relatively small quantum circuit possesses greater descriptive power than a large neural network. Identifying specific applications where training can be accelerated using quantum computation is crucial for both reinforcement learning and quantum computing.

\bibliography{sn-bibliography}


\begin{thebibliography}{36}
\ifx \bisbn   \undefined \def \bisbn  #1{ISBN #1}\fi
\ifx \binits  \undefined \def \binits#1{#1}\fi
\ifx \bauthor  \undefined \def \bauthor#1{#1}\fi
\ifx \batitle  \undefined \def \batitle#1{#1}\fi
\ifx \bjtitle  \undefined \def \bjtitle#1{#1}\fi
\ifx \bvolume  \undefined \def \bvolume#1{\textbf{#1}}\fi
\ifx \byear  \undefined \def \byear#1{#1}\fi
\ifx \bissue  \undefined \def \bissue#1{#1}\fi
\ifx \bfpage  \undefined \def \bfpage#1{#1}\fi
\ifx \blpage  \undefined \def \blpage #1{#1}\fi
\ifx \burl  \undefined \def \burl#1{\textsf{#1}}\fi
\ifx \doiurl  \undefined \def \doiurl#1{\url{https://doi.org/#1}}\fi
\ifx \betal  \undefined \def \betal{\textit{et al.}}\fi
\ifx \binstitute  \undefined \def \binstitute#1{#1}\fi
\ifx \binstitutionaled  \undefined \def \binstitutionaled#1{#1}\fi
\ifx \bctitle  \undefined \def \bctitle#1{#1}\fi
\ifx \beditor  \undefined \def \beditor#1{#1}\fi
\ifx \bpublisher  \undefined \def \bpublisher#1{#1}\fi
\ifx \bbtitle  \undefined \def \bbtitle#1{#1}\fi
\ifx \bedition  \undefined \def \bedition#1{#1}\fi
\ifx \bseriesno  \undefined \def \bseriesno#1{#1}\fi
\ifx \blocation  \undefined \def \blocation#1{#1}\fi
\ifx \bsertitle  \undefined \def \bsertitle#1{#1}\fi
\ifx \bsnm \undefined \def \bsnm#1{#1}\fi
\ifx \bsuffix \undefined \def \bsuffix#1{#1}\fi
\ifx \bparticle \undefined \def \bparticle#1{#1}\fi
\ifx \barticle \undefined \def \barticle#1{#1}\fi
\bibcommenthead
\ifx \bconfdate \undefined \def \bconfdate #1{#1}\fi
\ifx \botherref \undefined \def \botherref #1{#1}\fi
\ifx \url \undefined \def \url#1{\textsf{#1}}\fi
\ifx \bchapter \undefined \def \bchapter#1{#1}\fi
\ifx \bbook \undefined \def \bbook#1{#1}\fi
\ifx \bcomment \undefined \def \bcomment#1{#1}\fi
\ifx \oauthor \undefined \def \oauthor#1{#1}\fi
\ifx \citeauthoryear \undefined \def \citeauthoryear#1{#1}\fi
\ifx \endbibitem  \undefined \def \endbibitem {}\fi
\ifx \bconflocation  \undefined \def \bconflocation#1{#1}\fi
\ifx \arxivurl  \undefined \def \arxivurl#1{\textsf{#1}}\fi
\csname PreBibitemsHook\endcsname

\bibitem[\protect\citeauthoryear{Zhu and Zhao}{2021}]{Zhu_2022}
\begin{barticle}
\bauthor{\bsnm{Zhu}, \binits{Z.}},
\bauthor{\bsnm{Zhao}, \binits{H.}}:
\batitle{A survey of deep rl and il for autonomous driving policy learning}.
\bjtitle{IEEE Transactions on Intelligent Transportation Systems}
\bvolume{23}(\bissue{9}),
\bfpage{14043}--\blpage{14065}
(\byear{2021})
\end{barticle}
\endbibitem

\bibitem[\protect\citeauthoryear{Chib and Singh}{2023}]{chib2023recent}
\begin{botherref}
\oauthor{\bsnm{Chib}, \binits{P.S.}},
\oauthor{\bsnm{Singh}, \binits{P.}}:
Recent advancements in end-to-end autonomous driving using deep learning: A survey.
IEEE Transactions on Intelligent Vehicles,
1--18
(2023)
\doiurl{10.1109/TIV.2023.3318070}
\end{botherref}
\endbibitem

\bibitem[\protect\citeauthoryear{Kiran et~al.}{2021}]{RL_Survey3}
\begin{barticle}
\bauthor{\bsnm{Kiran}, \binits{B.R.}},
\bauthor{\bsnm{Sobh}, \binits{I.}},
\bauthor{\bsnm{Talpaert}, \binits{V.}},
\bauthor{\bsnm{Mannion}, \binits{P.}},
\bauthor{\bsnm{Al~Sallab}, \binits{A.A.}},
\bauthor{\bsnm{Yogamani}, \binits{S.}},
\bauthor{\bsnm{P{\'e}rez}, \binits{P.}}:
\batitle{Deep reinforcement learning for autonomous driving: A survey}.
\bjtitle{IEEE Transactions on Intelligent Transportation Systems}
\bvolume{23}(\bissue{6}),
\bfpage{4909}--\blpage{4926}
(\byear{2021})
\end{barticle}
\endbibitem

\bibitem[\protect\citeauthoryear{Pusse and Klusch}{2019}]{hyleap}
\begin{bchapter}
\bauthor{\bsnm{Pusse}, \binits{F.}},
\bauthor{\bsnm{Klusch}, \binits{M.}}:
\bctitle{Hybrid online pomdp planning and deep reinforcement learning for safer self-driving cars}.
In: \bbtitle{2019 IEEE Intelligent Vehicles Symposium (IV)},
pp. \bfpage{1013}--\blpage{1020}
(\byear{2019}).
\doiurl{10.1109/IVS.2019.8814125}
\end{bchapter}
\endbibitem

\bibitem[\protect\citeauthoryear{Gupta and Klusch}{2023}]{gupta2022hybrid}
\begin{bchapter}
\bauthor{\bsnm{Gupta}, \binits{D.}},
\bauthor{\bsnm{Klusch}, \binits{M.}}:
\bctitle{Hylear: Hybrid deep reinforcement learning and planning for safe and comfortable automated driving}.
In: \bbtitle{2023 IEEE Intelligent Vehicles Symposium (IV)},
pp. \bfpage{1}--\blpage{8}
(\byear{2023}).
\doiurl{10.1109/IV55152.2023.10186781}
\end{bchapter}
\endbibitem

\bibitem[\protect\citeauthoryear{Jerbi et~al.}{2021}]{jerbi2021parametrized}
\begin{barticle}
\bauthor{\bsnm{Jerbi}, \binits{S.}},
\bauthor{\bsnm{Gyurik}, \binits{C.}},
\bauthor{\bsnm{Marshall}, \binits{S.}},
\bauthor{\bsnm{Briegel}, \binits{H.}},
\bauthor{\bsnm{Dunjko}, \binits{V.}}:
\batitle{Parametrized quantum policies for reinforcement learning}.
\bjtitle{Advances in Neural Information Processing Systems}
\bvolume{34},
\bfpage{28362}--\blpage{28375}
(\byear{2021})
\end{barticle}
\endbibitem

\bibitem[\protect\citeauthoryear{Lan}{2021}]{lan2021variational}
\begin{botherref}
\oauthor{\bsnm{Lan}, \binits{Q.}}:
Variational quantum soft actor-critic.
arXiv preprint:2112.11921
(2021)
\end{botherref}
\endbibitem

\bibitem[\protect\citeauthoryear{Chen et~al.}{2020}]{chen2020variational}
\begin{barticle}
\bauthor{\bsnm{Chen}, \binits{S.Y.-C.}},
\bauthor{\bsnm{Yang}, \binits{C.-H.H.}},
\bauthor{\bsnm{Qi}, \binits{J.}},
\bauthor{\bsnm{Chen}, \binits{P.-Y.}},
\bauthor{\bsnm{Ma}, \binits{X.}},
\bauthor{\bsnm{Goan}, \binits{H.-S.}}:
\batitle{Variational quantum circuits for deep reinforcement learning}.
\bjtitle{IEEE Access}
\bvolume{8},
\bfpage{141007}--\blpage{141024}
(\byear{2020})
\end{barticle}
\endbibitem

\bibitem[\protect\citeauthoryear{Skolik et~al.}{2022}]{Skolik_2022}
\begin{barticle}
\bauthor{\bsnm{Skolik}, \binits{A.}},
\bauthor{\bsnm{Jerbi}, \binits{S.}},
\bauthor{\bsnm{Dunjko}, \binits{V.}}:
\batitle{Quantum agents in the gym: a variational quantum algorithm for deep q-learning}.
\bjtitle{Quantum}
\bvolume{6},
\bfpage{720}
(\byear{2022})
\end{barticle}
\endbibitem

\bibitem[\protect\citeauthoryear{Brockman et~al.}{2016}]{brockman2016openai}
\begin{botherref}
\oauthor{\bsnm{Brockman}, \binits{G.}},
\oauthor{\bsnm{Cheung}, \binits{V.}},
\oauthor{\bsnm{Pettersson}, \binits{L.}},
\oauthor{\bsnm{Schneider}, \binits{J.}},
\oauthor{\bsnm{Schulman}, \binits{J.}},
\oauthor{\bsnm{Tang}, \binits{J.}},
\oauthor{\bsnm{Zaremba}, \binits{W.}}:
Openai gym.
arXiv preprint:1606.01540
(2016)
\end{botherref}
\endbibitem

\bibitem[\protect\citeauthoryear{P{\'e}rez-Salinas et~al.}{2020}]{P_rez_Salinas_2020}
\begin{barticle}
\bauthor{\bsnm{P{\'e}rez-Salinas}, \binits{A.}},
\bauthor{\bsnm{Cervera-Lierta}, \binits{A.}},
\bauthor{\bsnm{Gil-Fuster}, \binits{E.}},
\bauthor{\bsnm{Latorre}, \binits{J.I.}}:
\batitle{Data re-uploading for a universal quantum classifier}.
\bjtitle{Quantum}
\bvolume{4},
\bfpage{226}
(\byear{2020})
\end{barticle}
\endbibitem

\bibitem[\protect\citeauthoryear{Mnih et~al.}{2016}]{A2C}
\begin{bchapter}
\bauthor{\bsnm{Mnih}, \binits{V.}},
\bauthor{\bsnm{Badia}, \binits{A.P.}},
\bauthor{\bsnm{Mirza}, \binits{M.}},
\bauthor{\bsnm{Graves}, \binits{A.}},
\bauthor{\bsnm{Lillicrap}, \binits{T.}},
\bauthor{\bsnm{Harley}, \binits{T.}},
\bauthor{\bsnm{Silver}, \binits{D.}},
\bauthor{\bsnm{Kavukcuoglu}, \binits{K.}}:
\bctitle{Asynchronous methods for deep reinforcement learning}.
In: \bbtitle{International Conference on Machine Learning},
pp. \bfpage{1928}--\blpage{1937}
(\byear{2016}).
\bcomment{PMLR}
\end{bchapter}
\endbibitem

\bibitem[\protect\citeauthoryear{Mirowski et~al.}{2016}]{mirowski2017learning}
\begin{botherref}
\oauthor{\bsnm{Mirowski}, \binits{P.}},
\oauthor{\bsnm{Pascanu}, \binits{R.}},
\oauthor{\bsnm{Viola}, \binits{F.}},
\oauthor{\bsnm{Soyer}, \binits{H.}},
\oauthor{\bsnm{Ballard}, \binits{A.J.}},
\oauthor{\bsnm{Banino}, \binits{A.}},
\oauthor{\bsnm{Denil}, \binits{M.}},
\oauthor{\bsnm{Goroshin}, \binits{R.}},
\oauthor{\bsnm{Sifre}, \binits{L.}},
\oauthor{\bsnm{Kavukcuoglu}, \binits{K.}}, et al.:
Learning to navigate in complex environments.
arXiv preprint:1611.03673
(2016)
\end{botherref}
\endbibitem

\bibitem[\protect\citeauthoryear{Everett et~al.}{2021}]{Everett_2021}
\begin{barticle}
\bauthor{\bsnm{Everett}, \binits{M.}},
\bauthor{\bsnm{Chen}, \binits{Y.F.}},
\bauthor{\bsnm{How}, \binits{J.P.}}:
\batitle{Collision avoidance in pedestrian-rich environments with deep reinforcement learning}.
\bjtitle{IEEE Access}
\bvolume{9},
\bfpage{10357}--\blpage{10377}
(\byear{2021})
\end{barticle}
\endbibitem

\bibitem[\protect\citeauthoryear{Hochreiter and Schmidhuber}{1997}]{LSTM}
\begin{barticle}
\bauthor{\bsnm{Hochreiter}, \binits{S.}},
\bauthor{\bsnm{Schmidhuber}, \binits{J.}}:
\batitle{{Long Short-Term Memory}}.
\bjtitle{Neural Computation}
\bvolume{9}(\bissue{8}),
\bfpage{1735}--\blpage{1780}
(\byear{1997})
\doiurl{10.1162/neco.1997.9.8.1735}
\end{barticle}
\endbibitem

\bibitem[\protect\citeauthoryear{Schulman et~al.}{2017}]{schulman2017proximal}
\begin{botherref}
\oauthor{\bsnm{Schulman}, \binits{J.}},
\oauthor{\bsnm{Wolski}, \binits{F.}},
\oauthor{\bsnm{Dhariwal}, \binits{P.}},
\oauthor{\bsnm{Radford}, \binits{A.}},
\oauthor{\bsnm{Klimov}, \binits{O.}}:
Proximal policy optimization algorithms.
arXiv preprint:1707.06347
(2017)
\end{botherref}
\endbibitem

\bibitem[\protect\citeauthoryear{Tang}{2019}]{tang2020learning}
\begin{bchapter}
\bauthor{\bsnm{Tang}, \binits{Y.}}:
\bctitle{Towards learning multi-agent negotiations via self-play}.
In: \bbtitle{Proceedings of the IEEE/CVF International Conference on Computer Vision Workshops}
(\byear{2019})
\end{bchapter}
\endbibitem

\bibitem[\protect\citeauthoryear{Bergholm et~al.}{2018}]{bergholm2022pennylane}
\begin{botherref}
\oauthor{\bsnm{Bergholm}, \binits{V.}},
\oauthor{\bsnm{Izaac}, \binits{J.}},
\oauthor{\bsnm{Schuld}, \binits{M.}},
\oauthor{\bsnm{Gogolin}, \binits{C.}},
\oauthor{\bsnm{Ahmed}, \binits{S.}},
\oauthor{\bsnm{Ajith}, \binits{V.}},
\oauthor{\bsnm{Alam}, \binits{M.S.}},
\oauthor{\bsnm{Alonso-Linaje}, \binits{G.}},
\oauthor{\bsnm{AkashNarayanan}, \binits{B.}},
\oauthor{\bsnm{Asadi}, \binits{A.}}, et al.:
Pennylane: Automatic differentiation of hybrid quantum-classical computations.
arXiv preprint arXiv:1811.04968
(2018)
\end{botherref}
\endbibitem

\bibitem[\protect\citeauthoryear{Lockwood and Si}{2020}]{lockwood2020reinforcement}
\begin{bchapter}
\bauthor{\bsnm{Lockwood}, \binits{O.}},
\bauthor{\bsnm{Si}, \binits{M.}}:
\bctitle{Reinforcement learning with quantum variational circuit}.
In: \bbtitle{Proceedings of the AAAI Conference on Artificial Intelligence and Interactive Digital Entertainment},
vol. \bseriesno{16},
pp. \bfpage{245}--\blpage{251}
(\byear{2020})
\end{bchapter}
\endbibitem

\bibitem[\protect\citeauthoryear{Sutton et~al.}{1999}]{reinforce}
\begin{botherref}
\oauthor{\bsnm{Sutton}, \binits{R.S.}},
\oauthor{\bsnm{McAllester}, \binits{D.}},
\oauthor{\bsnm{Singh}, \binits{S.}},
\oauthor{\bsnm{Mansour}, \binits{Y.}}:
Policy gradient methods for reinforcement learning with function approximation.
Advances in neural information processing systems
\textbf{12}
(1999)
\end{botherref}
\endbibitem

\bibitem[\protect\citeauthoryear{Geva and Sitte}{1993}]{cartpole}
\begin{barticle}
\bauthor{\bsnm{Geva}, \binits{S.}},
\bauthor{\bsnm{Sitte}, \binits{J.}}:
\batitle{A cartpole experiment benchmark for trainable controllers}.
\bjtitle{Control Systems, IEEE}
\bvolume{13},
\bfpage{40}--\blpage{51}
(\byear{1993})
\doiurl{10.1109/37.236324}
\end{barticle}
\endbibitem

\bibitem[\protect\citeauthoryear{Kwak et~al.}{2021}]{kwak2021introduction}
\begin{bchapter}
\bauthor{\bsnm{Kwak}, \binits{Y.}},
\bauthor{\bsnm{Yun}, \binits{W.J.}},
\bauthor{\bsnm{Jung}, \binits{S.}},
\bauthor{\bsnm{Kim}, \binits{J.-K.}},
\bauthor{\bsnm{Kim}, \binits{J.}}:
\bctitle{Introduction to quantum reinforcement learning: Theory and pennylane-based implementation}.
In: \bbtitle{2021 International Conference on Information and Communication Technology Convergence (ICTC)},
pp. \bfpage{416}--\blpage{420}
(\byear{2021}).
\bcomment{IEEE}
\end{bchapter}
\endbibitem

\bibitem[\protect\citeauthoryear{Acuto et~al.}{2022}]{acuto2022variational}
\begin{botherref}
\oauthor{\bsnm{Acuto}, \binits{A.}},
\oauthor{\bsnm{Barill{\`a}}, \binits{P.}},
\oauthor{\bsnm{Bozzolo}, \binits{L.}},
\oauthor{\bsnm{Conterno}, \binits{M.}},
\oauthor{\bsnm{Pavese}, \binits{M.}},
\oauthor{\bsnm{Policicchio}, \binits{A.}}:
Variational quantum soft actor-critic for robotic arm control.
arXiv:2212.11681
(2022)
\end{botherref}
\endbibitem

\bibitem[\protect\citeauthoryear{Kandala et~al.}{2017}]{Kandala_2017}
\begin{barticle}
\bauthor{\bsnm{Kandala}, \binits{A.}},
\bauthor{\bsnm{Mezzacapo}, \binits{A.}},
\bauthor{\bsnm{Temme}, \binits{K.}},
\bauthor{\bsnm{Takita}, \binits{M.}},
\bauthor{\bsnm{Brink}, \binits{M.}},
\bauthor{\bsnm{Chow}, \binits{J.M.}},
\bauthor{\bsnm{Gambetta}, \binits{J.M.}}:
\batitle{Hardware-efficient variational quantum eigensolver for small molecules and quantum magnets}.
\bjtitle{nature}
\bvolume{549}(\bissue{7671}),
\bfpage{242}--\blpage{246}
(\byear{2017})
\end{barticle}
\endbibitem

\bibitem[\protect\citeauthoryear{Dosovitskiy et~al.}{2017}]{dosovitskiy2017carla}
\begin{bchapter}
\bauthor{\bsnm{Dosovitskiy}, \binits{A.}},
\bauthor{\bsnm{Ros}, \binits{G.}},
\bauthor{\bsnm{Codevilla}, \binits{F.}},
\bauthor{\bsnm{Lopez}, \binits{A.}},
\bauthor{\bsnm{Koltun}, \binits{V.}}:
\bctitle{Carla: An open urban driving simulator}.
In: \bbtitle{Conference on Robot Learning},
pp. \bfpage{1}--\blpage{16}
(\byear{2017}).
\bcomment{PMLR}
\end{bchapter}
\endbibitem

\bibitem[\protect\citeauthoryear{Bartels and Erbsmehl}{2014}]{GIDAS}
\begin{botherref}
\oauthor{\bsnm{Bartels}, \binits{B.}},
\oauthor{\bsnm{Erbsmehl}, \binits{C.}}:
Bewegungsverhalten von fu{\ss}g{\"a}ngern im stra{\ss}enverkehr, teil 1.
FAT-Schriftenreihe
(267)
(2014)
\end{botherref}
\endbibitem

\bibitem[\protect\citeauthoryear{Ba et~al.}{2016}]{ba2016layer}
\begin{botherref}
\oauthor{\bsnm{Ba}, \binits{J.L.}},
\oauthor{\bsnm{Kiros}, \binits{J.R.}},
\oauthor{\bsnm{Hinton}, \binits{G.E.}}:
Layer normalization.
arXiv preprint arXiv:1607.06450
(2016)
\end{botherref}
\endbibitem

\bibitem[\protect\citeauthoryear{Mitarai et~al.}{2018}]{Mitarai_2018}
\begin{barticle}
\bauthor{\bsnm{Mitarai}, \binits{K.}},
\bauthor{\bsnm{Negoro}, \binits{M.}},
\bauthor{\bsnm{Kitagawa}, \binits{M.}},
\bauthor{\bsnm{Fujii}, \binits{K.}}:
\batitle{Quantum circuit learning}.
\bjtitle{Physical Review A}
\bvolume{98}(\bissue{3}),
\bfpage{032309}
(\byear{2018})
\end{barticle}
\endbibitem

\bibitem[\protect\citeauthoryear{Schuld et~al.}{2019}]{Schuld_2019}
\begin{barticle}
\bauthor{\bsnm{Schuld}, \binits{M.}},
\bauthor{\bsnm{Bergholm}, \binits{V.}},
\bauthor{\bsnm{Gogolin}, \binits{C.}},
\bauthor{\bsnm{Izaac}, \binits{J.}},
\bauthor{\bsnm{Killoran}, \binits{N.}}:
\batitle{Evaluating analytic gradients on quantum hardware}.
\bjtitle{Physical Review A}
\bvolume{99}(\bissue{3}),
\bfpage{032331}
(\byear{2019})
\end{barticle}
\endbibitem

\bibitem[\protect\citeauthoryear{{Rigetti Computing}}{Accessed: 2023-09-19}]{pyquil-noise}
\begin{botherref}
\oauthor{\bsnm{{Rigetti Computing}}}:
Quantum Gate Errors - PyQuil Documentation.
Website.
\url{https://pyquil-docs.rigetti.com/en/stable/noise.html#quantum-gate-errors}
\end{botherref}
\endbibitem

\bibitem[\protect\citeauthoryear{King}{2003}]{king2002capacity}
\begin{barticle}
\bauthor{\bsnm{King}, \binits{C.}}:
\batitle{The capacity of the quantum depolarizing channel}.
\bjtitle{IEEE Transactions on Information Theory}
\bvolume{49}(\bissue{1}),
\bfpage{221}--\blpage{229}
(\byear{2003})
\end{barticle}
\endbibitem

\bibitem[\protect\citeauthoryear{Berezniuk et~al.}{2020}]{berezniuk2001scale}
\begin{botherref}
\oauthor{\bsnm{Berezniuk}, \binits{O.}},
\oauthor{\bsnm{Figalli}, \binits{A.}},
\oauthor{\bsnm{Ghigliazza}, \binits{R.}},
\oauthor{\bsnm{Musaelian}, \binits{K.}}:
A scale-dependent notion of effective dimension.
arXiv preprint arXiv:2001.10872
(2020)
\end{botherref}
\endbibitem

\bibitem[\protect\citeauthoryear{Abbas et~al.}{2021}]{Abbas_2021}
\begin{barticle}
\bauthor{\bsnm{Abbas}, \binits{A.}},
\bauthor{\bsnm{Sutter}, \binits{D.}},
\bauthor{\bsnm{Zoufal}, \binits{C.}},
\bauthor{\bsnm{Lucchi}, \binits{A.}},
\bauthor{\bsnm{Figalli}, \binits{A.}},
\bauthor{\bsnm{Woerner}, \binits{S.}}:
\batitle{The power of quantum neural networks}.
\bjtitle{Nature Computational Science}
\bvolume{1}(\bissue{6}),
\bfpage{403}--\blpage{409}
(\byear{2021})
\end{barticle}
\endbibitem

\bibitem[\protect\citeauthoryear{Grant et~al.}{2019}]{Grant_2019}
\begin{barticle}
\bauthor{\bsnm{Grant}, \binits{E.}},
\bauthor{\bsnm{Wossnig}, \binits{L.}},
\bauthor{\bsnm{Ostaszewski}, \binits{M.}},
\bauthor{\bsnm{Benedetti}, \binits{M.}}:
\batitle{An initialization strategy for addressing barren plateaus in parametrized quantum circuits}.
\bjtitle{Quantum}
\bvolume{3},
\bfpage{214}
(\byear{2019})
\end{barticle}
\endbibitem

\bibitem[\protect\citeauthoryear{Skolik et~al.}{2021}]{Skolik_2021}
\begin{barticle}
\bauthor{\bsnm{Skolik}, \binits{A.}},
\bauthor{\bsnm{McClean}, \binits{J.R.}},
\bauthor{\bsnm{Mohseni}, \binits{M.}},
\bauthor{\bsnm{Smagt}, \binits{P.}},
\bauthor{\bsnm{Leib}, \binits{M.}}:
\batitle{Layerwise learning for quantum neural networks}.
\bjtitle{Quantum Machine Intelligence}
\bvolume{3},
\bfpage{1}--\blpage{11}
(\byear{2021})
\end{barticle}
\endbibitem

\bibitem[\protect\citeauthoryear{Holmes et~al.}{2022}]{Holmes_2022}
\begin{barticle}
\bauthor{\bsnm{Holmes}, \binits{Z.}},
\bauthor{\bsnm{Sharma}, \binits{K.}},
\bauthor{\bsnm{Cerezo}, \binits{M.}},
\bauthor{\bsnm{Coles}, \binits{P.J.}}:
\batitle{Connecting ansatz expressibility to gradient magnitudes and barren plateaus}.
\bjtitle{PRX Quantum}
\bvolume{3}(\bissue{1}),
\bfpage{010313}
(\byear{2022})
\end{barticle}
\endbibitem

\end{thebibliography}

\section*{Code and Data Availability}

All code to generate the data, figures and analyses in this study is publicly available with detailed information on the implementation via the following repository: \href{https://github.com/roboak/Nav-Q}{https://github.com/roboak/Nav-Q}.

\end{document}